\documentclass{article}
\usepackage{a4wide}
\usepackage{latexsym}
\usepackage[english]{babel}
\usepackage{amsmath,amssymb}
\usepackage{amsfonts}
\usepackage{xspace,ifthen}
\usepackage{tikz}
\usepackage{listings}
\usepackage{todonotes}
\usepackage{hyperref}
\definecolor{codegreen}{rgb}{0,0.6,0}
\definecolor{codegray}{rgb}{0.5,0.5,0.5}
\definecolor{codepurple}{rgb}{0.58,0,0.82}
\definecolor{backcolour}{rgb}{0.96,0.96,0.96}
\definecolor{darkbrown}{rgb}{0.00,0.00,0.50}

\lstdefinelanguage{mCRL2}
{ morecomment=[l]{\%},
  keywords={sort, cons, map, var, eqn, struct, act, proc, init, sum, forall, exists, true, false, mu, nu, val}
}  
\lstdefinelanguage{tools}
{ morecomment=[l]{\%},
  keywords={mcrl22lps, lps2lts, lpsconstelm, lpsparelm, lpsfununfold, lpsparunfold, lpssuminst, pbesparelm, pbesconstelm,
            pbessolvesymbolic, pbesrewr, pbesstategraph, lpsstategraph, lpsrewr, lps2pbes}
}

\lstdefinestyle{mystyle}{
    backgroundcolor=\color{backcolour},   
    commentstyle=\color{codegreen},
    keywordstyle=\color{magenta},
    numberstyle=\tiny\color{codegray},
    stringstyle=\color{codepurple},
    basicstyle=\ttfamily\footnotesize,
    identifierstyle=\color{darkbrown},
    breakatwhitespace=false,         
    breaklines=true,                 
    captionpos=b,                    
    keepspaces=true,                 
    numbers=left,                    
    numbersep=5pt,                  
    showspaces=false,                
    showstringspaces=false,
    showtabs=false,                  
    tabsize=2
}

\newcounter{theoremcnt}[section]
\renewcommand{\thetheoremcnt}{\thesection.\arabic{theoremcnt}}

{\begin{trivlist}\refstepcounter{theoremcnt}
\item{\bf Example \thetheoremcnt.}}
{\end{trivlist}}

{\begin{trivlist}\refstepcounter{theoremcnt}
\item{\bf Definition \thetheoremcnt.}}
{\end{trivlist}}

{\begin{trivlist}\refstepcounter{theoremcnt}
\item{\bf Remark \thetheoremcnt.}}
{\end{trivlist}}

\newenvironment{definition-arg}[1]%
{\begin{trivlist}\refstepcounter{theoremcnt}
\item{\bf Definition \thetheoremcnt\ (#1).}}
{\end{trivlist}}

{\begin{trivlist}\refstepcounter{theoremcnt}
\item{\bf Lemma \thetheoremcnt.}}
{\end{trivlist}}

\newenvironment{lemma-arg}[1]%
{\begin{trivlist}\refstepcounter{theoremcnt}
\item{\bf Lemma \thetheoremcnt\ (#1).}}
{\end{trivlist}}

{\begin{trivlist}\refstepcounter{theoremcnt}
\item{\bf Theorem \thetheoremcnt.}}
{\end{trivlist}}

\newenvironment{theorem-arg}[1]%
{\begin{trivlist}\refstepcounter{theoremcnt}
\item{\bf Theorem \thetheoremcnt\ (#1).}}
{\end{trivlist}}

{\begin{trivlist}\refstepcounter{theoremcnt}
\item{\bf Exercise \thetheoremcnt.}}{\end{trivlist}}

\newenvironment{exercise*}%
{\begin{trivlist}\refstepcounter{theoremcnt}
\item{\bf $\bigstar$Exercise \thetheoremcnt.}}{\end{trivlist}}

{\begin{trivlist} \item{\bf Proof.}}{~\hfill$\quad\Box$\end{trivlist}}

\newcounter{requirementcnt}[subsection]
\renewcommand{\therequirementcnt}{\thesubsection.\arabic{requirementcnt}}

\newenvironment{requirement}[2]
{\begin{trivlist}\refstepcounter{requirementcnt}\label{#1}
\item{\textbf{Requirement \therequirementcnt: #2}}\\ \noindent}{\vspace{-1.5ex}\end{trivlist}}

\newcommand{\Bool}{\mathbb{B}}

\newcommand{\sembracks}[1]{\ensuremath{|\!|#1|\!|}}
\newcommand{\sem}[2][]{\ensuremath{\ifthenelse{\equal{#1}{}}{\sembracks{#2}}{\sembracks{#2}{#1}}}}

\newcommand{\formula}[1]{$\texttt{\textcolor{darkbrown}{#1}}$}

\title{\textsf{A formal specification of the desired software behaviour\\ of the
Princess Marijke lock complex}}
\author{Jan Friso Groote, Matthias Volk\footnote{The research has been supported by the Interreg North Sea STORM\_SAFE project.\\\includegraphics{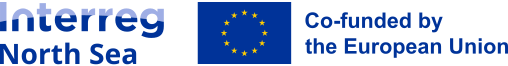}}\\\
\small{Eindhoven University of Technology, The Netherlands}\\
\small\texttt{\{J.F.Groote,M.Volk\}@tue.nl}}
\date{}

\begin{document}
\maketitle
\begin{abstract}
\noindent%
The Princess Marijke lock complex is a large lock and water-protection installation in the Netherlands 
between the river Rhine and the Amsterdam-Rijnkanaal---a large waterway 
connecting the Rhine to the port of Amsterdam. The lock complex consists of 
two independent locks and a moveable flood-protection barrier.
Ensuring safe control of the lock complex is of utmost importance to guarantee both flood-protection and reliable ship operations.

This paper gives a precise, formal description of the software control of the lock complex in less than 400 lines of mCRL2 code.
This description can act as a blueprint on how the software of this lock complex needs to be constructed. 
Moreover, using model checking, 53 software requirements are shown to be valid, 
ensuring that the formal description of the behaviour is correct with regard to these
properties and is unlikely to contain mistakes and oversights. 
\end{abstract}

\section{Introduction}
Infrastructural objects, such as locks, bridges and tunnels are increasingly
computer controlled. This has numerous advantages. The software allows
the objects to operate with increased autonomy while
automatically checking for safe and prompt operation. 

The extensive use of computers has also a substantial downside. Often it is not
known how the software behaves precisely, especially when it has been provided by commercial parties without an adequate
description of how the software is supposed to behave. As generally 
the concrete software is large, hard to read, and often completely inaccessible,
it is generally impossible to extract the precise behaviour from the software itself. 

For normal day-to-day behaviour of such control software this is often not so much an issue,
as malfunction is easily detected by testing, and repaired. But it is another story
for rarer behaviour such as dealing with sensor or actuator failures, or less common normal operation. 
In those cases the software can behave in unexpected and undesired ways, with problems ranging from 
annoyances, such as incorrect messages to the operator, unpleasant behaviour, such as bridges failing to 
open or close \cite{Haringvlietbrug2025}, to outright dangerous situations such as bridges opening 
without barriers going down \cite{ketelbrug2009}, or 
doors of flood gates that are open during high tide \cite{IJmuiden2023}. 
As the size and complexity of the controlling software continues to grow, such issues will increase in number,
unless we design and document the expected behaviour of this software very carefully and build the software
according to such detailed specifications. This means that 
the desired software needs to be formally specified, 
and this specification needs to be precisely analysed to avoid potential flaws and mishaps. 

This situation can be compared to the `hardware' of such artefacts
where it is completely inconceivable that an infrastructural
object is being built without having a precise and mathematically analysed blueprint of how it will look like. 
Already in ancient times technical drawings were used to assist and illustrate construction. 
From the renaissance onwards technical drawings became more commonplace. Orthographic projections
were used from the 18th century, the 19th century saw systematic use and education, whereas
the 20th century saw systematic standardisation and the assistance by computer
\cite{booker1963history,rovida2012machines}. The introduction of 
these engineering drawings also facilitated the development of techniques to evaluate
the strength and durability of mechanical designs before construction. 

When it comes to software controlled systems there is no accepted standard 
or systematic practice in designing and documenting the software control of
these systems. Documentation of software, if it exists, often consists of textual descriptions, sometimes
supported with some semi-formal drawing techniques for behaviour such as SysML~\cite{SysML2015} or UML~\cite{UMLspec}, 
or even low level code~\cite{DSO}. Generally, such descriptions are not very precise,
although they are often very extensive. As such, these descriptions are a mixed blessing if it comes to its 
value for documentation and construction of the actual software, and they are hardly usable to mathematically
evaluate the design of the software. 

In this paper we show -- what we believe -- is a proper way of of describing such software, namely by
using a precise and concise formal description, in this case in mCRL2~\cite{DBLP:books/mit/GrooteM2014}. 
Concretely, we specify the control of the \textit{Princess Marijke lock complex} located near the city Wijk bij Duurstede in the Netherlands, and operated by Rijkswaterstaat, the Dutch agency for public works and water management.
The lock complex connects the river Rhine to the Amsterdam-Rijnkanaal.
The latter is a human-constructed waterway between the Rhine and the port of Amsterdam.
The Princess Marijke lock complex consists of two locks and one flood barrier. Under normal circumstances the barrier is open and the locks are not used.
In case of high water in the Rhine, the barrier is closed to protect the hinterland, and the locks are
used to let commercial shipping proceed. 

The locks and the flood barrier are controlled by a lock controller.
The controller reacts on commands from a human operator and controls the infrastructure elements, for instance opening or closing the lock gates and flood barrier, setting the traffic lights or changing the water level.
We provide a precise specification of the behaviour of the lock controller in mCRL2.
In particular, our specification explicitly incorporates potential failures of components, for example lights that do not show the right aspect, motors that fail or gates that can get stuck. 

We base our work on public information, mainly~\cite{DBLP:conf/ccta/ReijnenVMR19}. In~\cite{DBLP:conf/ccta/ReijnenVMR19} 
a description is provided of the infrastructure of the Princess Marijke lock complex, called the plant, 
requirements are listed, and a controller is automatically generated for the plant satisfying
the requirements. Contrary to our work, the description in~\cite{DBLP:conf/ccta/ReijnenVMR19}
does not include the option that components fail, such as 
traffic lights being broken.
We did intentionally not discuss our specification extensively with experts responsible
for the operation of the locks, to avoid disclosing information that they may not want to
become public. As such this model may deviate from the actually desired behaviour of the locks. 

But from our modelling and a few discussions with the experts, 
we believe that it is not fixed what the `desired behaviour' of the locks is.
This is due to the fact
that they look at lock complexes from different perspectives than modelling the behaviour of the controller.
The following example illustrates this. In the national standard describing bridges and locks~\cite{LBS_stopinstructie} it is prescribed that the traffic
lights governing the entry into one of the locks must go to a red colour whenever a lock goes into emergency mode.
The emergency mode is caused by an operator pressing an emergency button. 
The question now is whether lights can be switched back to green while the lock is in emergency mode. 
We asked this to experts, but received as only response that this question is indeed interesting and relevant.
In our model we choose to allow the traffic lights to go to green while in emergency mode as this was done
in the CIF model~\cite{DBLP:conf/ccta/ReijnenVMR19}, but
this feels unnatural, and we believe this requires further discussion. There are more places with different options 
for the design of the ideal behaviour, and in such cases, we simply took reasonable choices. 
Strangely enough, the description in~\cite{DBLP:conf/ccta/ReijnenVMR19} does not satisfy the prescription in~\cite{LBS_stopinstructie}
and does not set the traffic lights to red while in an emergency. But note that due to the formally precise
description in~\cite{DBLP:conf/ccta/ReijnenVMR19} we could determine without any doubt that there are such
deviations in behaviour, and this precision is positive. 

As we base our work mainly on public information 
we cannot claim that we made \textit{the} behavioural model of the software control of the Princess Marijke lock complex. 
This however does not diminish the value of our specification which is a fully reasonable description
of desired software for a lock complex, is completely precise, and consists of
less than 400 lines of mCRL2 code. The full specification is provided in Appendix~\ref{app:model}. This might be compared to the slightly less
than 3000 lines of specification belonging to~\cite{DBLP:conf/ccta/ReijnenVMR19},
which does not include the generated model of the controller.
 
The conciseness of our model does not say that its design was easy. Not only did we encounter design questions
which we in some cases resolved arbitrarily, but we also made our fair share of specifications mistakes, some
at a conceptual level, and some as simple mistypings and copy and paste errors. 

To counter such errors, it is essential to check separately formulated behavioural requirements on the 
specification \cite{DBLP:journals/scp/BrandG15}.
We verified 53 software requirements, checking both safety and liveness aspects, leading to various
adaptations of the model. These requirements have been formalised in the modal mu-calculus~\cite{DBLP:books/mit/GrooteM2014},
and provided in Appendix~\ref{app:formulas}. To our surprise, some formulas became quite extensive,
where in one case the modal formula requires more than 200 lines. This is mainly due to the incorporation of failure behaviour.

The necessity of double checking requirements is 
nicely illustrated in~\cite{DBLP:conf/ccta/ReijnenVMR19}. Here requirements are formulated and the controller
is automatically generated out of the requirements. At least one essential
safety requirement was not specified. This led to the situation where their generated controller could start
engines while in emergency mode. This omission was not detected in their simulation and development environment
and only became known when we pointed it out.

Although the modal formulas form the larger part
of this paper, they are just supportive, and the essential specification of the software
are the 374 lines of mCRL2 behavioural specification.

\paragraph{Outline}
This paper has the following structure. In Chapter~\ref{ch::marijkelock} an outline of the Princess Marijke lock
complex is given. In Chapter~\ref{ch::formalcontroller} the model of the controller is explained. 
We do not explain the formal modelling language of mCRL2, for which we refer to sources such as~\cite{DBLP:books/mit/GrooteM2014, DBLP:books/sp/AtifG23, DBLP:conf/facs2/GrooteKLVW19, DBLP:conf/forte/GrooteK21}.
The model itself is provided in Appendix~\ref{app:model}. In the rather lengthy Chapter~\ref{sec:requirements} 
the behavioural requirements are listed and their verification is reported. The translation of the
requirements to the modal mu-calculus 
is given in Appendix~\ref{app:formulas}. A short conclusion is provided in Chapter~\ref{ch::conclusion}. 

\paragraph{Acknowledgement}
We thank Jochem Smit for pinpointing an error in the specification of the Princess Marijke lock complex 
and Johan van den Bogaard for general comments.

\section{The Princess Marijke lock complex}
\label{ch::marijkelock}
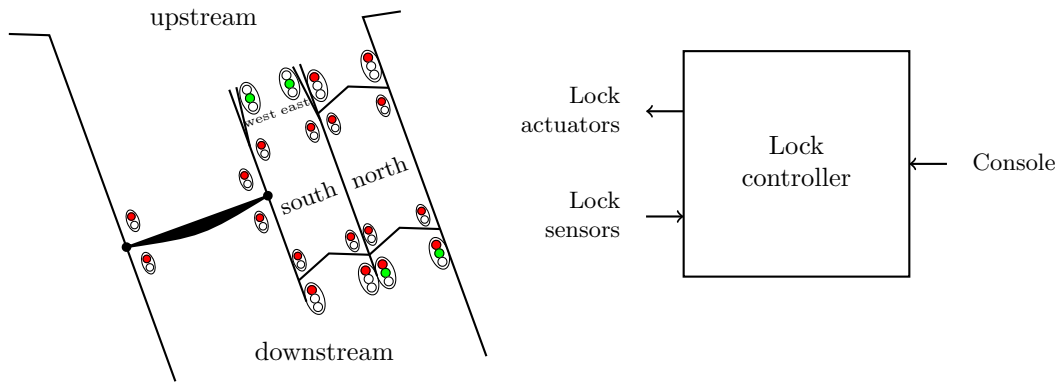
\begin{figure}[t]
\begin{center}
\begin{tikzpicture}[auto,rotate=20,transform shape]{c}
\draw [thick] (1,0.6) -- (1,5.5) -- (0.5,5.7);
\draw [thick] (3,1) -- (3,4);
\draw [thick] (4,1) -- (4,4);
\draw [thick] (5,-0.5) -- (5,4.3) -- (5.5,4.2);
\draw [thick,  fill] (1,2.5) circle(0.05) -- (3,2.5) circle(0.05) .. controls (2,2.3) .. (1,2.5);
\draw [thick] (3,1.3) -- (3.5, 1.5) -- (4,1.3);
\draw [thick] (4,1.3) -- (4.5, 1.5) -- (5,1.3);
\draw [thick] (3,3.2) -- (3.1, 4.0) (3.9,4.0) -- (4,3.2);
\draw [thick] (4,3.3) -- (4.5, 3.5) -- (5,3.3);
\draw (3,0.3) node[rotate=-20] {$\text{downstream}$};
\draw (3,5) node[rotate=-20] {$\text{upstream}$};
\draw (4.5,2.3) node {$\text{\small{north}}$};
\draw (3.5,2.3) node {$\text{\small south}$};
\draw (3.25,3.5) node {$\text{\tiny west}$};
\draw (3.75,3.5) node {$\text{\tiny east}$};
\newcommand{\simplelight}[4]{
\draw (#1,#2) ellipse(0.08 and 0.15);
\draw [fill, #3](#1,#2+0.06) circle(0.05);
\draw (#1,#2+0.06) circle(0.05);
\draw [fill, #4](#1,#2-0.06) circle(0.05);
\draw (#1,#2-0.06) circle(0.05);
}
\newcommand{\complexlight}[5]{
\draw (#1,#2) ellipse(0.12 and 0.22);
\draw [fill, #3](#1,#2+0.12) circle(0.06);
\draw (#1,#2+0.12) circle(0.06);
\draw [fill, #4](#1,#2) circle(0.06);
\draw (#1,#2) circle(0.06);
\draw [fill, #5](#1,#2-0.12) circle(0.06);
\draw (#1,#2-0.12) circle(0.06);
}
\simplelight{1.2}{2.2}{red}{white}
\simplelight{2.8}{2.2}{red}{white}
\simplelight{1.2}{2.8}{red}{white}
\simplelight{2.8}{2.8}{red}{white}
\simplelight{3.1}{1.55}{red}{white}
\simplelight{3.85}{1.55}{red}{white}
\simplelight{3.15}{3.1}{red}{white}
\simplelight{3.85}{3.1}{red}{white}

\complexlight{3.12}{1.0}{red}{white}{white}
\complexlight{3.88}{1.0}{red}{white}{white}
\complexlight{3.22}{3.8}{white}{green}{white}
\complexlight{3.78}{3.8}{white}{green}{white}

\simplelight{4.1}{1.55}{red}{white}
\simplelight{4.85}{1.55}{red}{white}
\simplelight{4.15}{3.1}{red}{white}
\simplelight{4.85}{3.1}{red}{white}

\complexlight{4.12}{1.0}{red}{green}{white}
\complexlight{4.88}{1.0}{red}{green}{white}
\complexlight{4.12}{3.65}{red}{white}{white}
\complexlight{4.88}{3.65}{red}{white}{white}
\end{tikzpicture}\vspace{-0.7cm}
\begin{tikzpicture}
\draw [thick] (9,1) -- +(0,3) -- +(3,3) -- +(3,0) -- cycle +(1.5,1.5) node {\begin{tabular}{c}Lock\\controller\end{tabular}};
\draw [thick, <-] (12,2.5) -- +(0.5,0) node[right] {\small\begin{tabular}{l}Console\end{tabular}};
\draw [thick, ->] (9,3.2) -- +(-0.5,0) node[left] {\small\begin{tabular}{r}Lock\\actuators\end{tabular}};
\draw [thick, <-] (9,1.8) -- +(-0.5,0) node[left] {\small\begin{tabular}{r}Lock\\sensors\end{tabular}};
\draw [white] (10,-0.3) circle(0.1); 

\end{tikzpicture}
\end{center}

\caption{The Princess Marijke lock complex and its controller}
\label{PrincessMarijkeSluis}
\end{figure}

\begin{figure}[b]
\begin{center}
\begin{tikzpicture}
\newcommand{\simplelight}[4]{
\draw (#1,#2) ellipse(0.21 and 0.38);
\draw [fill, #3](#1,#2+0.17) circle(0.15);
\draw (#1,#2+0.17) circle(0.15);
\draw [fill, #4](#1,#2-0.17) circle(0.15);
\draw (#1,#2-0.17) circle(0.15);
}
\newcommand{\complexlight}[5]{
\draw (#1,#2) ellipse(0.25 and 0.57);
\draw [fill, #3](#1,#2+0.32) circle(0.15);
\draw (#1,#2+0.32) circle(0.15);
\draw [fill, #4](#1,#2) circle(0.15);
\draw (#1,#2) circle(0.15);
\draw [fill, #5](#1,#2-0.32) circle(0.15);
\draw (#1,#2-0.32) circle(0.15);
}
\simplelight{0}{1.2}{red}{white}
\draw [right] (0.3,1.2) node{\begin{tabular}{l}Forbidden to pass.\end{tabular}};
\simplelight{0}{0}{white}{green}
\draw [right] (0.3,0) node{\begin{tabular}{l}Allowed to pass.\end{tabular}};
\complexlight{5}{1.2}{red}{white}{white}
\draw [right] (5.3,1.2) node{\begin{tabular}{l}Forbidden to pass;\\lock is being operated.\end{tabular}};
\complexlight{5}{0}{red}{white}{red}
\draw [right] (5.3,0) node{\begin{tabular}{l}Forbidden to pass;\\ lock is not being operated.\end{tabular}};
\complexlight{10.5}{1.2}{red}{green}{white}
\draw [right] (10.8,1.2) node{\begin{tabular}{l}Forbidden to pass;\\ passage will be allowed\\shortly.\end{tabular}};
\complexlight{10.5}{0}{white}{green}{white}
\draw [right] (10.8,0) node{\begin{tabular}{l}Allowed to pass.\end{tabular}};
\end{tikzpicture}
\end{center}

\caption{The aspects of the single traffic light (left) and double traffic light (middle/right)}
\label{TrafficLights}
\end{figure}
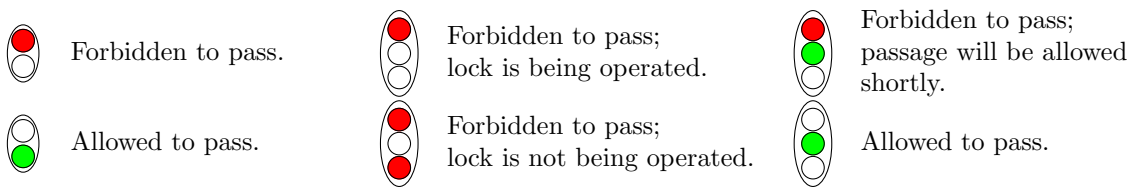

The Princess Marijke lock complex consists of two independent locks and a water barrier, see Figure~\ref{PrincessMarijkeSluis} at the left. 
The lock complex is oriented in the northwestern direction and this is why we depict it in a rotated fashion. 
The side connecting to the river Rhine is called the \textit{upstream} side and the side connecting to the Amsterdam-Rijnkanaal is
referred to as \textit{downstream}. Under normal operation, when the Rhine and the Amsterdam-Rijnkanaal have the same
water level, the barrier at the left is open and ship traffic can pass unhindered. 

\paragraph{Infrastructure elements}
When the water level of the Rhine and the Amsterdam-Rijnkanaal differ, the barrier is closed and the ships must use the locks.
As this is one of the major waterways in the Netherlands, the Princess Marijke lock complex has two separate locks, called \textit{south}
and \textit{north}. At both the upstream and the downstream side of each lock there are two gates, one referred to as the gate 
at the \textit{east} side and one gate at the \textit{west} side. The gates can be opened, closed and stopped when necessary. 
Each gate has a paddle that can be opened, closed and stopped
to let water into or out of the lock to level the water height in the lock
with that outside the lock. 
Each gate and paddle has a sensor to determine whether
the gate or paddle is fully closed, fully opened or somewhere in between. For each pair of gates there is also a sensor to determine
whether the water at both sides of the gate is at the same level or not. All sensors can fail. 

\paragraph{Traffic lights}
As can be seen in Figure~\ref{PrincessMarijkeSluis} there are a large number of traffic lights.
A \textit{single traffic light} can show only one light at the time, either red or green, for which it uses two separate lights. 
Single traffic lights are also called \textit{leaving traffic lights} and indicate whether a ship can leave a lock, or whether traversal through the water barrier is allowed.
A \textit{double traffic light} can show either one or two lights, namely single red or green, red-red or red-green, 
for which it uses three actual lights.
Double traffic lights are also called \textit{entering traffic lights} and indicate whether a ship can enter a lock.
In Figure~\ref{TrafficLights} the meaning of the different traffic light aspects is explained. 
Each light has a sensor to determine whether the light is working and which aspect it is currently
showing. Due to malfunction, the sensor of each traffic light can indicate another colour than the traffic light was instructed to show. 
The sensor can also indicate failure, not providing information on the actual aspect. 

\paragraph{Controller}
The lock complex is operated by an operator, who can either be present locally at the locks, or act remotely from
a central operating command centre. The operator can open, close and stop the barrier, the gates, and the paddles, 
and set all traffic lights.
Due to symmetry, the traffic lights, gates and paddles at the east and west side in a lock are operated simultaneously. 
The software controller permits the operator to open and close gates, paddles and barriers when it is safe to do so. 
For instance, the doors can only move when the lights show red. 
There is one exception, namely the barrier can close when its traffic lights are set to red, but may show another
aspect or even fail. This is to prevent that the barrier cannot be closed when its lights malfunction, avoiding
the risk of flooding of the hinterland due to the failure of a single light bulb. 

There are three emergency buttons, one for each lock and one for the barrier. When an emergency is activated, all movable 
parts in the relevant
lock or the barrier should come to a standstill and all traffic lights should go to red if they are green or 
red-green~\cite{LBS_stopinstructie}. 
During the emergency no movable part can be activated until the emergency is 
deactivated. It is possible to change the 
traffic lights while in emergency mode. In particular, they can be changed back to green before ending the emergency status. 

To operate the locks safely and adequately, there are a number of rules that the software controller enforces.
Here we give a number of examples. In Section~\ref{sec:requirements} an extensive list of such requirements
is given, which are verified on the model of the software controller.
\begin{itemize}
\item
The gates of a lock can only open if the gates and paddles at the opposite side are closed, the traffic lights
of that gate do not show green and the water at both sides of the gate is at the same level.
\item
When the movement of the gates of a lock is stopped, and the entering traffic lights show red-green, the traffic lights 
are switched to red~\cite{LBS_stopinstructie}.
\item
The traffic lights of the barrier and the locks will show red or red-red when the lock or barrier goes into
an emergency~\cite{LBS_stopinstructie}.
\item
The barrier can at any moment be closed, by giving the commands to terminate the emergency state, set the traffic lights to 
red and terminate any emergency status, provided the software can instruct all actuators, and read all
sensors, although the concrete responses of the sensors do not matter. 
\end{itemize}

Note that the model described in~\cite{DBLP:conf/ccta/ReijnenVMR19} deviates from some of the rules above and also does
not deal with failing behaviour. The model as we provide here very precisely prescribes a possible and reasonable 
behaviour of the software controller which is guaranteed to satisfy all requirements formulated in Section \ref{sec:requirements}.

\section{The formal specification of the software controller}
\label{ch::formalcontroller}
In this section we describe a model in mCRL2~\cite{DBLP:books/mit/GrooteM2014} of the behaviour of the software 
controller of the Princess Marijke lock complex. 
This model consists of a precise, but abstract description of all potential input and output behaviour of the controller. 
As such it describes all behaviour that the controller can ever execute. 
The controller receives commands and sensor information from the lock complex, and sends out commands to its actuators. 
The full mCRL2 specification is given in Appendix~\ref{app:model} and we refer in this section to lines in this specification.
As stated before, our model is mainly based on the description in~\cite{DBLP:conf/ccta/ReijnenVMR19}, and has explicitly not 
been discussed in detail with Rijkswaterstaat. 

The model consists of multiple parts.
First, the model describes the data types that form the contents of the messages that the controller exchanges.
These data types are given in Section~\ref{sec:model_data_types}.
Second, the messages which are sent and received are given in the form of abstract actions as described in Section~\ref{sec:model_input_output}.
Third, the behaviour model in Section~\ref{sec:model_behaviour} describes all possible sequences of such actions, exactly capturing all the potential behaviour that the controller can exhibit.

\subsection{Data types}
\label{sec:model_data_types}
\begin{table}[t]
\begin{center}
\begin{tabular}{|l|p{3.0cm}|p{7.4cm}|}
\hline
Data type &Elements & Description\\
\hline
$\mathit{Lock}$&$\mathit{north},\mathit{south}$&The identifiers of the locks.\\
$\mathit{StreamSide}$&$\mathit{upstream}$, $\mathit{downstream}$&The river side, resp.\ the waterway side of the lock.\\
$\mathit{Orientation}$&$\mathit{east},\mathit{west}$&Identifiers for the right and left side of the lock.\\
\hline
$\mathit{ConsoleCommand}$&$\mathit{command\_open}$, $\mathit{command\_close}$, $\mathit{command\_stop}$&Instructions from the console or an external source to
the lock controller.\\
$\mathit{EmergencyCommand}$&$\mathit{activate},\mathit{deactivate}$&Emergency command instructions.\\
$\mathit{ActuatorCommand}$&$\mathit{do\_open},\,\,$$\mathit{do\_close}$, $\mathit{do\_emergencyStop}$, 
$\mathit{do\_endStopClosing}$, $\mathit{do\_endStopOpening}$&
Instructions to the barrier, gates and paddles to open, close and stop.\\
$\mathit{SensorPosition}$&$\mathit{sense\_open}$, $\mathit{sense\_closed}$, $\mathit{sense\_intermediate}$, $\mathit{fail\_position}$&The data that sensors in the barrier, gates and paddles can provide. It can be open, closed or in between or
the sensor can fail. \\
$\mathit{SingleLight}$&$\mathit{red},\mathit{green}$&The possible aspects of a single light.\\
$\mathit{DoubleLight}$&$\mathit{single\_red}$, $\mathit{single\_green}$, $\mathit{redred},\mathit{redgreen}$&The possible aspects of a double light.\\
$\mathit{SingleLightStatus}$&$\mathit{show}(\mathit{red})$, $\mathit{show}(\mathit{green})$, $\mathit{fail\_single}$&
	   Indications from a single light sensor, which can be either of type $\mathit{SingleLight}$ or a sensor failure.\\
$\mathit{DoubleLightStatus}$&$\mathit{show}(\mathit{single\_red})$, $\mathit{show}(\mathit{single\_green})$, 
                              $\mathit{show}(\mathit{redred})$, $\mathit{show}(\mathit{redgreen})$, $\mathit{fail\_double}$&
	   Indications from a double light sensor, which can be either of type $\mathit{DoubleLight}$ or a sensor failure.\\
$\mathit{WaterLevel}$&$\mathit{equal},\,\,$$\mathit{unequal}$, $\mathit{fail\_water\_sensor}$&Output of the water sensor, which can be equal, not equal, or fail.\\

\hline
\end{tabular}
\end{center}
\caption{The data types used in the communications of the controller of the Princess Marijke lock complex}
\label{table:datatypes}
\end{table}
The data types of the data that the controller exchanges with its environment are given in Table~\ref{table:datatypes}
and are described at lines 1--20 of the specification in Appendix~\ref{app:model}. Using 
the keyword $\texttt{\textcolor{magenta}{sort}}$ new data types are declared in a style common in functional programming. 
Note that the data types indicate which elements exist in each data type, but that it does not specify in any sense
how these data types are implemented. 

The data types can be divided into two groups. The first group consists of
$\mathit{Lock}$, $\mathit{StreamSide}$ and $\mathit{Orientation}$.
These are indicators of the infrastructure as depicted in Figure~\ref{PrincessMarijkeSluis} and are giving the names 
$\mathit{north}$ and $\mathit{south}$
of the locks, $\mathit{upstream}$ and $\mathit{downstream}$ as stream sides, 
and $\mathit{east}$ and $\mathit{west}$ for orientations 
within each lock. 

The second group consists of the commands or sensor information that is communicated by or to the controller. 
One example, $\mathit{ActuatorCommand}$, contains the instructions to open, close or stop a barrier, gate or paddle. 
Other examples regard the setting and reading of the traffic lights. 

\subsection{Inputs and outputs of the controller}
\label{sec:model_input_output}
\begin{table}
\begin{center}
\begin{tabular}{|p{0.39\textwidth}|p{0.55\textwidth}|}
\hline
$\textit{GateCommand}(l,s,c)$&Instruction to execute console command $c$ on the gates in lock $l$ at stream side $s$.\\
$\textit{PaddleCommand}(l,s,c)$&Instruction to execute console command $c$ on the paddles in lock $l$ at stream side $s$.\\
$\textit{EmergencyLockCommand}(l,c)$&Instruction to execute emergency command $c$ on lock~$l$.\\
$\textit{BarrierCommand}(c)$&Instruction to execute console command $c$ on the barrier.\\
$\textit{EmergencyBarrierCommand}(c)$&Instruction to execute emergency command $c$ on the barrier.\\
$\textit{EnteringTrafficLightCommand}(l,s,c)$&Instruction to set the entering traffic lights in lock $l$ at stream side $s$ to double colour $c$.\\
$\textit{LeavingTrafficLightCommand}(l,s,c)$&Instruction to set the leaving traffic lights in lock $l$ at stream side $s$ to single colour $c$.\\
$\textit{BarrierTrafficLightCommand}(s,c)$&Instruction to set the traffic lights of the barrier at stream side $s$ to single colour $c$.\\
\hline
\end{tabular}
\end{center}
\caption{The external inputs for the controller of the Princess Marijke lock complex}
\label{tab:inputs}
\end{table}
\begin{table}
\begin{center}
\begin{tabular}{|p{0.39\textwidth}|p{0.55\textwidth}|}
\hline
$\textit{GateActuator}(l,s,o,c)$&Instruct the gate in lock $l$ at stream side $s$ and orientation $o$ to execute output command $c$.\\
$\textit{PaddleActuator}(l,s,o,c)$&Instruct the paddle in lock $l$ at stream side $s$ and orientation $o$ to execute output command $c$.\\
$\textit{BarrierActuator}(c)$&Instruct the barrier to execute output command $c$.\\
$\textit{EnteringTrafficLightActuator}(l,s,o,c)$&Set the entering traffic light for lock $l$ at stream side $s$ and orientation $o$ to double colour $c$.\\
$\textit{LeavingTrafficLightActuator}(l,s,o,c)$&Set the leaving traffic light for lock $l$ at stream side $s$ and orientation $o$ to single colour $c$.\\ 
$\textit{BarrierTrafficLightActuator}(s,o,c)$&Set the traffic light of the barrier at stream side $s$ and orientation $o$ to single colour $c$.\\
\hline
\end{tabular}
\end{center}
\caption{The actuator commands from the controller to the Princess Marijke lock complex}
\label{tab:actuator_commands}
\end{table}

\begin{table}
\begin{center}
\begin{tabular}{|p{0.39\textwidth}|p{0.55\textwidth}|}
\hline
$\textit{GateSensor}(l,s,o,p)$&Receive sensor input $p$ of the gate in lock $l$ at stream side $s$ and orientation $o$.\\
$\textit{PaddleSensor}(l,s,o,p)$&Receive sensor input $p$ from the paddle in lock $l$ at stream side $s$ and orientation $o$.\\
$\textit{BarrierSensor}(p)$&Receive sensor information $p$ from the barrier.\\
$\textit{WaterSensor}(l,s,w)$&Receive the water height difference $w$ for lock $l$ at stream side $s$.\\
$\textit{EnteringTrafficLightSensor}(l,s,o,c)$&Receive the double light status $c$ of the entering traffic light in lock $l$ at stream side $s$ and orientation $o$. \\
$\textit{LeavingTrafficLightSensor}(l,s,o,c)$&Receive the single light status $c$ of the leaving traffic light in lock $l$ at stream side $s$ and orientation $o$. \\ 
$\textit{BarrierTrafficLightSensor}(s,o,c)$&Receive the single traffic light status $c$ of the barrier at stream side $s$ and orientation $o$.\\
\hline
\end{tabular}
\end{center}
\caption{The sensor inputs from the Princess Marijke lock complex to the controller}
\label{tab:sensors}
\end{table}
Tables~\ref{tab:inputs},~\ref{tab:actuator_commands} and~\ref{tab:sensors} abstractly describe the external commands to the controller, the instructions to actuators, and the inputs from sensors, respectively.
In the specification 
they are declared at lines 22--44 preceded by the keyword $\texttt{\textcolor{magenta}{act}}$. 
There is one extra action $\mathit{skip}$ that is an action that represents doing nothing, and it is only part of the
specification for technical reasons. 

When an action occurs in the specification, it indicates only that that action can take place. 
The actions do not describe how the action is implemented in the software. For instance, the action 
$\mathit{GateCommand}(\mathit{north},\mathit{upstream},\mathit{command\_open})$ indicates that the gates at the upstream
side of the northern lock must be opened. This functionality could be implemented in various ways, for example by sending a message via a network to the gate or by changing an electrical signal to the gate.

\subsection{The behaviour of the lock controller}
\label{sec:model_behaviour}
\begin{table}
\begin{center}
\begin{tabular}{|p{0.31\textwidth}|p{0.19\textwidth}|p{0.42\textwidth}|}
\hline
Data type &Elements & Description\\
\hline
$\small\mathit{Position}$&\small$\mathit{opening}, \mathit{closing}$, $\mathit{opened}$, $\mathit{closed}$&\small Possible positions of the barrier, gates and paddles.\\
$\small\textit{LockStreamSideOrientationTriple}$&\small triple $(l, s, o)$&\small Identifier of a lock $l$ at stream side $s$ and orientation $o$.\\
$\small\textit{LockStreamSideTuple}$&\small tuple $(l, s)$&\small Identifier of a lock $l$ at stream side $s$.\\
\hline
\end{tabular}
\end{center}
\caption{The data types of the lock controller}
\label{tab:data_types}
\end{table}
\begin{table}
\begin{center}
\begin{tabular}{|p{0.19\textwidth}|p{0.31\textwidth}|p{0.42\textwidth}|}
\hline
Parameter name&Type&Description\\
\hline
$\small\textit{BarrierStatus}$&$\small\textit{Position}$&\small Position of the barrier.\\
$\small\textit{BarrierInEmergency}$&$\Bool$&\small Indicates that the barrier is in emergency.\\
$\small\textit{BarrierLightStatus}$&$\small\textit{StreamSide}\rightarrow\textit{SingleLight}$&
\small Instructed colours of the barrier traffic lights.\\
$\small\textit{GateStatus}$&$\small\textit{LockStreamSideOrientationTriple}$ $\rightarrow\textit{Position}$&
\small The positions of the gates.\\
$\small\textit{PaddleStatus}$&$\small\textit{LockStreamSideOrientationTriple}$ $\rightarrow\textit{Position}$&
\small The positions of the paddles.\\
$\small\textit{EnteringLightStatus}$&\hspace{-1.7ex}\small\begin{tabular}[t]{l}$\textit{LockStreamSideTuple}\rightarrow$\\
$\textit{DoubleLight}$\end{tabular}&
\small The instructed aspects of the entering lights.\\
$\small\textit{LeavingLightStatus}$&\hspace{-1.7ex}\small\begin{tabular}[t]{l}$\textit{LockStreamSideTuple}\rightarrow$\\
$\textit{SingleLight}$\end{tabular}&
\small The instructed aspects of the leaving lights.\\
$\small\textit{WaterLevelStatus}$&\hspace{-1.7ex}\small\begin{tabular}[t]{l}$\textit{LockStreamSideTuple}\rightarrow$\\
$\textit{WaterLevel}$\end{tabular}&
\small The measured water levels.\\
$\small\textit{LocksInEmergency}$&$\small\textit{FSet}(\textit{Lock})$&\small Set with the locks that are in emergency.\\
\hline
\end{tabular}
\end{center}
\caption{The parameters of the lock controller}
\label{tab:parameters}
\end{table}

The behaviour of the software controller is given at lines 70--363 in the mCRL2 model in the form of a process $\mathit{Controller}$
which is declared using the keyword $\texttt{\textcolor{magenta}{proc}}$. 
The process has 9 parameters that form its internal state and store the status of the barrier, gates, lights, etc.
The parameters are described in Table~\ref{tab:parameters}.
The first column gives the names of the parameters and the third column gives a short description.
The second column provides the type of each parameter.
These types are declared at lines 47--52 and further described in Table~\ref{tab:data_types}.
The type $\Bool$ represents the booleans. A type of the shape $D\rightarrow E$
is a function from $D$ to $E$.
For instance, $\textit{GateStatus}$ is a function that maps each gate, located in a lock
at some stream side and some orientation, to its position, being open, closed, opening or closing.
When the position
of a gate is instructed to change, the $\textit{GateStatus}$ is updated.
The type $\textit{FSet}(\textit{Lock})$ is a finite set of locks. Whenever a lock is put in emergency mode, it is added to this set.
Note that for the traffic lights we combine both orientations, west and east, into a single parameter
as the traffic lights at both sides receive exactly the same instructions. 

After these type declarations some auxiliary functions are declared at lines 53--68 using the
keyword $\texttt{\textcolor{magenta}{map}}$. The functions are defined using the subsequent equations, preceded by $\texttt{\textcolor{magenta}{eqn}}$.
One example is function $\textit{opposite}$ which returns the opposite stream side.

From line 80 on, the behaviour of the controller is described. The specification of the behaviour follows a general structure: first an action with some inputs can happen, which then leads to subsequent input actions
checking values of sensors. Afterwards, some output actions instructing actuators take place and finally the parameters of the process are updated.
This structure typically looks as follows:
\begin{lstlisting}[language=mCRL2]
          sum l:Lock,s:StreamSide,o:Orientation. GateSensor(l,s,o,sense_open).
               (GateStatus(triple(l,s,o)) in { opening, opened })
                -> GateActuator(l,s,o, do_endStopOpening).
                   Controller(GateStatus=GateStatus[triple(l,s,o)->opened])
                <> Controller(GateStatus=GateStatus[triple(l,s,o)->closing])+
          ....
\end{lstlisting}

The \texttt{\textcolor{darkbrown}{+}} at the end is the choice operator indicating that either this behaviour can be executed, or the other behaviours indicated by the dots.
The $\texttt{\textcolor{magenta}{sum}}$ at the beginning says that
the behaviour can be done for an \texttt{\textcolor{darkbrown}{l}} of type \texttt{\textcolor{darkbrown}{Lock}},
\texttt{\textcolor{darkbrown}{s}} of type \texttt{\textcolor{darkbrown}{StreamSide}} and \texttt{\textcolor{darkbrown}{o}} of type \texttt{\textcolor{darkbrown}{Orientation}}. Hence, the action \texttt{\textcolor{darkbrown}{GateSensor(l,s,o,sense\_open)...}}
expresses that if \texttt{\textcolor{darkbrown}{sense\_open}} is sensed for any lock $l$, stream side $s$ and 
orientation $o$ then the subsequent behaviour at the dots is executed. 

The next operator is an if-then-else, denoted as \texttt{\textcolor{darkbrown}{condition -> then-part <> else-part}}. 
In the example it is expressed that if the condition \texttt{\textcolor{darkbrown}{GateStatus(triple(l,s,o))}} is either 
\texttt{\textcolor{darkbrown}{opening}} or 
\texttt{\textcolor{darkbrown}{opened}}, then the gate actuator for lock $l$, stream side $s$
and orientation $o$ receives the command \texttt{\textcolor{darkbrown}{do\_endStopOpening}}, and the main process
continues recording that this gate is now \texttt{\textcolor{darkbrown}{opened}}. Otherwise, no action is performed,
and it is recorded that the gate is \texttt{\textcolor{darkbrown}{closing}}. 

At line 366 the initial values of the parameters of the controller are specified, behind the keyword 
$\texttt{\textcolor{magenta}{init}}$. 

Below all the parts of the controller are explained separately. 

\subsubsection{Lock emergency commands}
At line 81 it is defined what happens if an activate emergency command for a lock is received and at line 114 it is
explained what happens if it receives a deactivate emergency command. 
In the latter case it is recorded that the lock is not in emergency status any more.

When the emergency mode of a lock is activated, all actuators of the gates and paddles receive an emergency stop instruction.
The traffic lights for the lock that show green are set to show one red light.
Traffic lights showing red remain as before.
Moreover, it is recorded
that traffic lights around the lock are set to red, and the lock is in emergency mode.

\subsubsection{Gate control}
Any pair of gates can be instructed to open (line 117), to close (line 144), or to stop (line 153). 

If a pair of gates is instructed to open, it is checked whether the opposite gates and paddles are closed,
the leaving and entering lights have not been instructed to show a single green light, the water level
around the gates is equalised, and the sensors of the traffic lights indicate that they 
are not actually showing a single green light or are failed.
If all checks are successful, then both gates are instructed to open,
and the process administration records that both gates are opening.
If the checks are unsuccessful, the gates cannot be opened and the lock remains as before.

If the controller receives the command to close a pair of gates (line 144), it is checked that the
traffic lights have been set to red and the lock is not in emergency mode. If so, the gates are instructed 
to close and it is administrated that these gates are now closing. Note that it is only checked that the lights
have been instructed to show red, but it is not checked
whether they actually show red. As closing the gates is safety critical, the operation cannot
depend on the correct functioning of the lights. 

When the controller is instructed to stop a pair of gates (line 153), an emergency stop is sent
to both gates and the traffic lights surrounding the gates are set to red. 

\subsubsection{Gate sensors}
The gate sensors can report on their status and at lines 171--188 it is shown how the controller
responds to this. If a gate is reported to be fully open (\texttt{\textcolor{darkbrown}{sense\_open}}) and it was opening or open, then it is stopped and it is administrated as being open.
If \texttt{\textcolor{darkbrown}{sense\_open}} is reported and the gate was closed or closing instead, then it is recorded as closing, as apparently it cannot be closed. The controller 
behaves similarly if a gate is reported to be closed.

At line 182 a gate reports that it is neither open nor closed, or it reports that its sensor
fails. In that case it is administrated that the gate is opening when it was opening or opened,
and that it is closing when it was closing or closed. 

\subsubsection{Paddle control}
The paddles can receive commands to open, close and stop (lines 189--213), similar to the gates. 
When the controller receives a command to open paddles, it checks whether the lock is not in emergency mode,
and that the opposite gates and paddles are closed. If so, it instructs the paddles to open.
The paddles in a lock can always close, unless the lock is in emergency mode. 
When an instruction by the controller is received to stop the paddles, this will always
immediately happen. 

\subsubsection{Paddle sensors}
Paddles can submit status information, namely whether they are open, closed, in an intermediate position
or whether a paddle sensor is malfunctioning (lines 214--231). The controller handles the paddle sensor in exactly the same way as the gate sensors. 

\subsubsection{Water sensors}
The handling of information from the water sensors is specified at lines 232--237. If a water level is reported
as equal, this is recorded. Unequal water levels, or messages on a failed sensor are recorded as an unequal level.

\subsubsection{Lock double traffic lights control}
In lines 238--270 it is indicated how console commands to set the double (entering) traffic lights are forwarded to
the traffic light actuators. Commands to set the traffic light to red or double red are always forwarded.
The entering traffic light can only be set to red-green if the leaving lights of that pair of gates are set to red.
The entering traffic lights can only be set to green if the corresponding leaving traffic lights are set to red and also show red, and the corresponding gates are open. 

\subsubsection{Lock single traffic lights control}
The control of the single (leaving) traffic lights for the locks is described at lines 271--292. 
A console command to set these traffic lights to red is always forwarded. The leaving traffic lights
can only be set to green if the entering traffic lights of the same pair of gates are set to single or double red 
and show red, and the corresponding gates are open. 

\subsubsection{Barrier emergency commands}
Lines 293--304 specify the behaviour for a barrier emergency.
If the barrier emergency is activated, the barrier is stopped, the barrier traffic lights are set to red, and it is 
recorded that the barrier is in emergency mode.
If the barrier emergency is deactivated, nothing happens, except that it is recorded that the barrier is not in emergency mode anymore. 
      
\subsubsection{Barrier control}
Lines 305--338 describe the handling of console commands to open, close and stop the barrier.
If a command is received to open the barrier, the barrier is not in emergency mode, and all traffic lights around the barrier are set to red and show red, then the barrier is instructed to open.

If a close command is received for the barrier, the barrier is not in emergency mode and the lights are instructed to go to red, then the close instruction is forwarded to the barrier actuator.
Note that for opening the barrier, it is checked whether the lights are red.
However, for closing, the lights are not checked, because the barrier must always be able to close, even if the barrier lights do not function properly.
               
Stop commands are always forwarded to the barrier. If the barrier receives and forwards a stop instruction,
all traffic lights are immediately set to red.

\subsubsection{Barrier traffic light}
At lines 339--349 it is indicated that the traffic lights of the barriers can always be set to red by the 
operator, and be set to green if the barrier is open. 

\subsubsection{Barrier sensor}
Lines 350--363 define how the controller responds if the barrier sends information about its status.
The behaviour is similar to the gate sensors.
If the barrier indicates that it is open, and it was supposed to be open or opening, then the
controller sends an instruction to stop opening the barrier, and records that the barrier is open.
If the barrier was not open or opening, it records that the barrier is closing.
This is especially important, when the barrier was supposed to be closed, but the controller receives the sensor value that it is open. In this case the status is set
to closing as apparently we cannot consider the barrier to be closed anymore.

If the barrier sends the sensor value that it is closed, the controller reacts similarly to the case when the barrier signalled that it was open. 
When the barrier sends an indication that the barrier is halfway, or that the sensor is malfunctioning, 
it is administrated to be opening if it was opened, closing if it was closed, and unchanged otherwise.

\section{Software requirements}
\label{sec:requirements}
It is difficult to know that a behavioural specification of the software controller is exactly as desired. 
The number of states of our controller is
$1.6 \cdot 10^{15}$ which shows the large number of different situations the controller has to handle. By checking explicit 
requirements on the software, the confidence is increased that the controller indeed performs as desired
\cite{DBLP:journals/scp/BrandG15}. Each requirement typically covers an aspect of the behaviour of the 
lock controller
and as such gives a different perspective on the behaviour than the mCRL2 specification, which is a complete
description of all potential behaviour of the lock controller. 
In this section the requirements are described textually.
In Appendix~\ref{app:formulas}, we translate the requirements into modal formulas, which precisely and concisely specify the desired behaviour.
All requirements have been shown to hold on the model. Checking the requirements led to the repair of a number of oversights in the model. 

In the following, we divide the requirements in four categories. 
\begin{itemize}
\item
\emph{Safety requirements} express that certain actions are only possible under particular circumstances. An example is that
gates of a lock can only be opened by the software controller if the lights are not green.
\item
\emph{Causality requirements} express that if an action happens, there is a good reason for it to happen. For instance, if
the software instructs the gates of a lock to open, a command to do so must have been received before. 
\item
\emph{Operator requirements} show that if an operator gives a command, the software controller gives
the desired instructions to the lock complex. For instance, if an activate emergency command is given for
a lock, the engines of the gates and paddles are stopped and all traffic lights of that lock are set to red. 
\item
\emph{Liveness requirements} express that something will happen under the right circumstances. For instance, 
if the instruction to open the gate of a lock is given to the software controller, it will instruct both gates
to open, provided the traffic lights do not show green, the water level at both sides of the gates is equal,
the opposing gates and paddles are closed and the lock is not in emergency mode. 
\end{itemize}

We were careful to cover all aspects of the lock with 53 requirements in total.
But it is not easy to know for sure that all requirements are part of our list. It is therefore useful to look at requirements
of similar objects. We benefited from the requirements in~\cite{DBLP:conf/ccta/ReijnenVMR19} by largely taking them
over. However, it is illustrative to note that the requirement `The barrier cannot move while in emergency 
mode' was forgotten in~\cite{DBLP:conf/ccta/ReijnenVMR19}. As the approach in~\cite{DBLP:conf/ccta/ReijnenVMR19} generates a model controller, and even its software, automatically, they failed to observe that this requirement was missing.
Their generated simulation model indeed allowed to start barrier engines in emergency mode and
this was only noted when we pointed it out.
Covering a large number of requirements, ideally created by independent teams, is therefore a good way to ensure that no important aspect was overlooked.

\subsection{Safety requirements}
Here we list the safety requirements of the Princess Marijke lock complex that together guarantee 
that the lock complex can only be operated in a safe way. 
\begin{requirement}{safreq1}{Opposing paddles cannot be both open simultaneously}
If the paddles are not closed at one stream side of a lock, the paddles at the other stream 
side cannot be opened. 
\end{requirement}
\begin{requirement}{safreq2}{Paddles cannot open with an opposing gate open}
If the gates are not closed at one stream side of a lock,
the paddles at the other stream side cannot be opened.
\end{requirement}
\begin{requirement}{safreq3}{Gates cannot open with an opposing paddle open}
If the paddles are not closed at one stream side of a lock, 
the gates at the other stream side cannot be opened.
\end{requirement}
\begin{requirement}{safreq4}{Gates cannot open with an opposing gate open}
If the gates are not closed at one stream side of a lock,
the gates at the other stream side cannot be opened.
\end{requirement}
\begin{requirement}{safreq5}{Gates can only open if the waterlevel is equal}
If the water level at a stream side of a lock is not level,
then both the east and west gates may not be opened.
\end{requirement}
\begin{requirement}{safreq6}{Traffic lights at entering and leaving side I}
If the traffic lights at the entering side of a lock are set to single green,
they must be red at the leaving side.
\end{requirement}
\begin{requirement}{safreq13}{Traffic lights at entering and leaving side II}
If the traffic lights at the leaving side of a lock are set to green,
they must be single red or red-red at the entering side. 
\end{requirement}
\begin{requirement}{safreq7}{Lights cannot be set to green if lock not open I}
If a gate is not measured to be open, the entering traffic lights cannot be
changed to green.
\end{requirement}
\begin{requirement}{safreq14}{Lights cannot be set to green if lock not open II}
If a gate is not measured to be open, the leaving traffic lights cannot be set to green.
\end{requirement}
\begin{requirement}{safreq8}{Gates cannot be closed if the lights are not set to red I}
If the entering traffic lights of a lock are not set to single red or red-red,
the gates cannot be instructed to close. 

We intentionally do not require that the lights are measured
to show single red or red-red, because if the lights fail, the operator should be allowed to close the gates.
Otherwise, protecting the hinterland may be prevented by the failure of a single traffic light to 
show the right aspect. 
\end{requirement}
\begin{requirement}{safreq9}{Gates cannot be closed if the lights are not set to red II}
If the leaving traffic lights are not set to red, the gates cannot be instructed to close. 
\end{requirement}
\begin{requirement}{safreq10}{Gates and paddles cannot move in emergency mode}
In emergency mode the gates and paddles actuators cannot be instructed to open or close.
\end{requirement}
\begin{requirement}{safreq23}{End stop opening gate only if open}
If a gate is instructed to do a \formula{do\_endStopOpening}, it is known to be open.
\end{requirement}
\begin{requirement}{safreq24}{End stop closing gate only if closed}
If a gate is instructed to do a \formula{do\_endStopClosing}, it is known to be closed. 
\end{requirement}
\begin{requirement}{safreq27}{End stop opening paddle only if open}
If a paddle is instructed to do a \formula{do\_endStopOpening}, it is known to be open. 
\end{requirement}
\begin{requirement}{safreq28}{End stop closing gate only if closed}
If a paddle is instructed to do a \formula{do\_endStopClosing}, it is known to be closed. 

\end{requirement}
\begin{requirement}{safreq29}{Barrier only closes when lights are red}
If the barrier actuator is instructed to close, then both the upstream and 
downstream traffic lights are set to red.

Note that it is explicitly not required that the traffic lights are measured
to be red. An operator should always be able to close the barrier to prevent flooding, even if the traffic lights fail, or
erroneously keep showing green despite being instructed to go to red. 
\end{requirement}
\begin{requirement}{safreq30}{Barrier lights only become green when the barrier is open}
If the barrier traffic lights are set to green, the barrier is open.
\end{requirement}
\begin{requirement}{safreq40}{The barrier cannot move in emergency mode}
In emergency mode the barrier cannot be instructed to open or close.
\end{requirement}
\begin{requirement}{safreq35}{End stop opening barrier only if the barrier is open}
If the barrier is instructed to do a \formula{do\_endStopOpening}, it is open. 
\end{requirement}
\begin{requirement}{safreq36}{End stop closing barrier only if the barrier is closed}
If the barrier is instructed to do an \formula{do\_endStopClosing}, it is closed. 
\end{requirement}
\subsection{Causality requirements}
The causality requirements say that no instruction to a device in the lock complex can be given without
good cause. For instance, a gate actuator can only be instructed to open if a console command to do so has
been issued after a previous instruction to that gate, see Requirement~\ref{causreq15} below. 

\begin{requirement}{causreq11}{Emergency stop of a gate}
If an emergency stop of a gate in a lock takes place, this has to be preceded 
by either an emergency stop of that lock, or the instruction to stop that
gate, which happened after the last instruction to that gate.
\end{requirement}
\begin{requirement}{causreq12}{Emergency stop of a paddle}
An emergency stop of a paddle in a lock has to be preceded 
by either an emergency stop of that lock, or the instruction to stop that
paddle, which happened after the last instruction to that paddle. 
\end{requirement}
\begin{requirement}{causreq15}{Opening a gate}
If a gate of a lock is opened, this has to be preceded by an instruction to open
that gate, and this instruction must have happened after the last command sent 
to the gate.
\end{requirement}
\begin{requirement}{causreq16}{Closing a gate}
If a gate of a lock is closed, this has to be preceded by an instruction to close that gate, and this instruction
must have happened after the last command sent to the gate. 
\end{requirement}
\begin{requirement}{causreq17}{Opening a paddle}
If a paddle of a lock is opened, this has to be preceded by an instruction to 
open that paddle, and this instruction must have happened after the last 
command sent to the paddle. 
\end{requirement}
\begin{requirement}{causreq18}{Closing a paddle}
If a paddle of a lock is closed, this has to be preceded by an instruction to 
close that paddle, and this instruction must have happened after the last 
command sent to the paddle. 
\end{requirement}
\begin{requirement}{causreq19}{Setting the entering lights in a lock}
If the traffic lights at the entering side of a lock are set to some colour,
an instruction to do so must have been given since the last instruction to the traffic lights.
In case the lights go to \formula{single\_red}, this can also be caused by activating emergency mode or
stopping a gate.
\end{requirement}
\begin{requirement}{causreq20}{Setting the leaving lights in a lock}
If the traffic lights at the leaving side of a lock are set to some colour,
an instruction to do so must have been given since the last instruction to the traffic lights.
In case the lights go to red, this can also be caused by activating an emergency.
\end{requirement}
\begin{requirement}{causreq31}{Setting the lights of the barrier}
If the traffic lights actuators at the barrier are set to green, a command to do so
must have been given. If the actuators are set to red, this must be due to an operator command to set them to red
or due to an emergency or due to a stop command for the barrier. 
\end{requirement}
\begin{requirement}{causreq32}{Opening the barrier}
If the barrier is opened, this has to be preceded by an instruction to open the barrier.
\end{requirement}
\begin{requirement}{causreq33}{Closing the barrier}
If the barrier is closed, this has to be preceded by an instruction to close the barrier.
\end{requirement}
\begin{requirement}{causreq34}{Stopping the barrier}
If an emergency stop of the barrier takes place, this has to be preceded by either an emergency stop of the barrier, 
or the instruction to stop that barrier.
\end{requirement}
\subsection{Operator requirements}
The operator requirements express that if the operator gives a particular instruction, 
this instruction is carried out,
provided the circumstances allow it. These requirements are typical examples of liveness properties, expressing
that particular behaviour will happen. 
\begin{requirement}{commandreq1}{Close command for the barrier}
If the barrier is instructed to close, it will close provided the traffic lights
show red and the barrier is not in emergency mode.
\end{requirement}
\begin{requirement}{commandreq2}{Open command for the barrier}
If the barrier is instructed to open, it will open provided the traffic
lights are set to red, the barrier is not in emergency mode, and
the traffic lights show red.
\end{requirement}
\begin{requirement}{commandreq3}{Stop command for the barrier}
If the barrier is instructed to stop, a stop command is sent to the actuator of the barrier, 
and all four traffic lights will be instructed to go to red.
\end{requirement}
\begin{requirement}{commandreq4}{Emergency command for the barrier}
If the barrier receives an emergency command, the engines are stopped
and all traffic lights of the barrier are switched to red. 
\end{requirement}
\begin{requirement}{commandreq5}{Lights command for the barrier}
If the barrier is instructed to show a certain aspect for its traffic lights
at a stream side, then this light is shown, except that for green the barrier must
be open. The barrier is open when it is sensed to be open after a valid 
open command has been given and the barrier did not receive another command afterwards.
An open command is valid if it is given when all lights are set to red and the
barrier is not in emergency mode, and it remains valid as long as all measurements of the lights
indicate that the lights show red. 

We like to point out that setting the lights is not influenced by the emergency mode. In particular,
in emergency mode the lights can be set to green. 

\end{requirement}
\begin{requirement}{commandreq6}{Close command for gates}
When the operator gives the command to close the gates,
the gates will be closed provided the lock is not in emergency mode
and the entering and leaving traffic lights are set to red.
\end{requirement}
\begin{requirement}{commandreq7}{Open command for gates}
When the operator gives the command to open the gates,
the gates will be opened provided the lock is not in emergency mode
and the entering and leaving traffic lights are set to red, and the opposite
gates and paddles are closed. 
\end{requirement}
\begin{requirement}{commandreq8}{Stop command for gates}
When the operator gives the command to stop the gates, the gates will
receive a stop command and the entering and leaving traffic lights are 
set to red.
\end{requirement}
\begin{requirement}{commandreq9}{Close command for paddles}
When the operator gives the command to close the paddles,
the paddles will be closed, if the lock is not in emergency mode.
\end{requirement}
\begin{requirement}{commandreq10}{Open command for paddles}
When the operator gives the command to open the paddles,
the paddles will open provided the lock is not in emergency mode
and the opposing gates and paddles are closed.
\end{requirement}
\begin{requirement}{commandreq11}{Stop command for paddles}
When the operator gives the command to stop the paddles, the paddles will
receive a stop command.
\end{requirement}
\begin{requirement}{commandreq12}{Emergency command for a lock}
If a lock receives an emergency instruction, the engines of the gates and paddles
are stopped and all entering and leaving traffic lights of the lock are switched to red
or red-red.
\end{requirement}
\begin{requirement}{commandreq13}{Leaving lights commands for a lock}
If the leaving lights are instructed to show a certain colour,
they will show this colour, except that green is only shown if
the gates are open. The gates are open if they were measured
to be open after having received a valid open command, which has
not been revoked. An opening command is valid if the opposite
paddles and gates are closed, the entering and leaving lights
are showing red, and the lock is not in emergency mode. An
opening command is revoked if the traffic lights around the
gates are measured not to show red, red-red or red-green.
\end{requirement}

\begin{requirement}{commandreq14}{Entering lights commands for a lock}
If the entering lights are instructed to show a certain colour,
they will show this colour, except in two cases. Red-green is only shown
if the opposite traffic light is red. Furthermore, single green is only shown if
the gates are open. The gates are open if they were measured
to be open after having received a valid open command, which has
not been revoked. An opening command is valid if the opposite
paddles and gates are closed, and the entering and leaving lights
are showing red, and the lock is not in emergency mode. An
opening command is revoked if the traffic lights around the
gates are measured not to show red, red-red or red-green.
\end{requirement}

\subsection{Liveness requirements}
We chose to formulate the liveness requirements to show that the essential functionalities
of the lock complex can be carried out. In a very much simplified form, 
these are formulated as follows: 
\begin{enumerate}
\item the barrier can always be closed, 
\item the gates in the locks can always be closed, and 
\item
a ship can always pass the locks. 
\end{enumerate}
Furthermore, three extra requirements are formulated that are explicitly stated in~\cite{LBS_stopinstructie}
expressing that traffic lights will go to red if gates are forced to stop, or the locks and the barrier go
into emergency mode.

\begin{requirement}{livereq1}{The barrier can always be closed}
At any time the barrier can always be closed by only performing the following actions. The operator
can set the traffic lights to red,
disable the emergency mode, and can instruct the barrier to close.
The controller is allowed to 
send instructions to the actuators for the gates, the paddles, the barrier and the lights. It can also
read the traffic light and water level sensors, where the actually sensed values do not matter. 
\end{requirement}
\begin{requirement}{livereq2}{Gates can always be closed}
At any time, it is possible to close a gate in a lock. For this the operator only needs to instruct the 
traffic lights of the lock to go to red, disable an emergency status of the lock, and give
the command to close the gates of the lock. In addition the controller is allowed to give actuator commands, 
and read the sensors, where the reported values of the sensors do not matter. 
\end{requirement}
\begin{requirement}{livereq3}{Ships can pass}
At any time, it is possible to let a ship pass. This is encoded by saying that
the actuators of the gates at both sides of a lock can be instructed
to start opening at any time. For this the operator can give the commands to open and close the gates, close
the paddles in the other lock, instruct the lights of that lock to go to
red, and by deactivating an emergency status of that lock. In addition the controller can give instructions all
actuators and read the sensors. For the sensors it is essential that the
lights correctly indicate that they are set to red, the sensors in the doors and
paddles of the lock must indicate that they are closed, and the water sensor
around the gates must report that the water level is equal. Opening the gates
does not depend on other values of the sensors being reported correctly. In 
order to deal properly with incorrect water measurements, we consider a 
water-equal measurement around a lock as valid if the paddle of the lock is open.
\end{requirement}
\begin{requirement}{livereq4}{Stopping gates prematurely}
When the gates are stopped opening or closing while the entering lights show green-red,
then the traffic lights will change to red. 
\end{requirement}
\begin{requirement}{livereq5}{Emergency stop of the lock}
When the emergency command is given for a lock, then the entering and leaving traffic lights will
change to red or red-red. 
\end{requirement}
\begin{requirement}{livereq6}{Emergency stop of the barrier}
When the emergency command is given for the barrier, then the traffic lights will change to red.
\end{requirement}

\subsection{Verification of the requirements}
The requirements where verified using the mCRL2 toolset~\cite{DBLP:conf/facs2/GrooteKLVW19,DBLP:books/mit/GrooteM2014},
were we especially relied on the symbolic solver for parity games~\cite{DBLP:conf/tacas/LaveauxWW22}. 
The verification was executed on a Mac Studio M1 Ultra with 128GB of memory with the development toolset of December 2024.
All 53 formulas were proven to hold on the model. The verification times varied
depending on the formula between 4 minutes and 31 hours. 

The verification procedure first linearised the behavioural mCRL2 specification into linear normal form. The resulting linear process was optimised
in various ways where in particular the function data types where replaced by a sequence of variables using the tool
\formula{lpsfununfold}. Subsequently, the optimised linear process and the requirement were 
transformed into a parameterised boolean equation system, which, after additional optimisations, was subsequently
solved using \formula{pbessolvesymbolic}.
The complete command sequence is provided in Appendix~\ref{app:script}.
The appendix also contains a script to generate the state space of the model.

\section{Conclusion}
\label{ch::conclusion}
We provided a precise, compact and complete specification of the behaviour of a controller of the Princess Marijke lock complex
where sensors and actuators can fail. The major purpose is to show that such behaviour can be denoted in an abstract
manner, while providing all essential behavioural details.
Because we included 
the option that sensors and actuators may fail, this behaviour is rather complex and versatile. 
There are various aspects where our choices in the behaviour can be debated and it is not unlikely that the
ultimate behavioural model will deviate from what we have described. 

We achieved high confidence that our model is free from unintended,
odd or erroneous behaviour by verifying a large set of behavioural requirements. 
While verifying, we found a number of issues in our model 
that we had to resolve, showing that verifying was absolutely necessary, 
fully in accordance with~\cite{DBLP:journals/scp/BrandG15}. 

We hope that our description is an inspiration for others on how to provide abstract and precise descriptions
of the control behaviour for large infrastructural objects. Note that there are other descriptions
of similarly modelled and verified infrastructural controllers, e.g., 
\cite{Beers24,LKCJG14,DBLP:journals/fac/BouwmanWLSR23,DBLP:journals/corr/abs-2403-18722}. 
We also hope and expect that such models will become the basis of more detailed mathematical analysis
of the behaviour of software controlled artefacts. For instance, there are proposals to determine
the risk of software failure based on formal models \cite{MartensProb24}, but also the influence of the 
software and the reliability of the sensors and actuators on the throughput of shipping can be determined
base on such models. 

\bibliographystyle{plain} 
\bibliography{references} 
\appendix
\section{Full mCRL2 model of the Princess Marijke lock complex}
\label{app:model}
In this appendix we give the full and precise mCRL2 model of the software controller.
In Section~\ref{ch::formalcontroller} this specification
is explained. 
\begin{lstlisting}[language=mCRL2]
% Infrastructure indicators.
sort Lock = struct north | south;
     StreamSide = struct upstream | downstream;
     Orientation = struct east | west;
% Command data structures.
     ConsoleCommand = struct command_open | command_close | command_stop;
     EmergencyCommand = struct activate | deactivate;
     ActuatorCommand = struct do_open | do_close | do_emergencyStop | 
                              do_endStopClosing | do_endStopOpening;
% Sensor data structures.
     SensorPosition = struct sense_open | sense_closed | 
                             sense_intermediate | fail_position;
% Colors of the lights. 
     SingleLight = struct red | green;
     DoubleLight = struct single_red | single_green | redred | redgreen; 
% Colors that the light sensors can give.
     SingleLightStatus = struct show(SingleLight) | fail_single; 
     DoubleLightStatus = struct show(DoubleLight) | fail_double;
% Waterheight indicators.
     WaterLevel = struct equal | unequal | fail_water_sensor;

% Operator commands.
act  GateCommand, PaddleCommand: Lock#StreamSide#ConsoleCommand;
     EmergencyLockCommand: Lock#EmergencyCommand;
     BarrierCommand: ConsoleCommand;
     EmergencyBarrierCommand: EmergencyCommand;
     EnteringTrafficLightCommand: Lock#StreamSide#DoubleLight;
     LeavingTrafficLightCommand: Lock#StreamSide#SingleLight;
     BarrierTrafficLightCommand: StreamSide#SingleLight;
% Actuator commands.
     GateActuator, PaddleActuator: Lock#StreamSide#Orientation#ActuatorCommand;
     BarrierActuator: ActuatorCommand;
     EnteringTrafficLightActuator: Lock#StreamSide#Orientation#DoubleLight;
     LeavingTrafficLightActuator: Lock#StreamSide#Orientation#SingleLight;
     BarrierTrafficLightActuator: StreamSide#Orientation#SingleLight;
% Sensor inputs.
     GateSensor, PaddleSensor: Lock#StreamSide#Orientation#SensorPosition;
     BarrierSensor: SensorPosition;
     WaterSensor: Lock#StreamSide#WaterLevel;
     EnteringTrafficLightSensor: Lock#StreamSide#Orientation#DoubleLightStatus;
     LeavingTrafficLightSensor: Lock#StreamSide#Orientation#SingleLightStatus;
     BarrierTrafficLightSensor: StreamSide#Orientation#SingleLightStatus;
% An action that represents doing nothing.
     skip;

% Data types used in the state of the lock controller. 
sort Position = struct opening | closing | opened | closed;
     LockStreamSideOrientationTriple = 
                        struct triple(lock: Lock, 
                                      streamside: StreamSide, 
                                      orientation: Orientation);
     LockStreamSideTuple = struct tuple(lock: Lock, streamside: StreamSide);
map  opposite: StreamSide -> StreamSide;
     startinternallockstatus, startinternalpaddlestatus: 
                                  LockStreamSideOrientationTriple -> Position;
     startinternalenteringtlstatus: LockStreamSideTuple -> DoubleLight;
     startinternalleavingtlstatus: LockStreamSideTuple -> SingleLight;
     startinternalbarriertlstatus: StreamSide -> SingleLight;
     startinternalwaterstatus: LockStreamSideTuple -> WaterLevel;
var  l:Lock, s:StreamSide, o:Orientation;
eqn  opposite(upstream) =  downstream;
     opposite(downstream) = upstream;
     startinternallockstatus(triple(l,s,o)) = opening;
     startinternalpaddlestatus(triple(l,s,o)) = opening;
     startinternalenteringtlstatus(tuple(l,s)) = single_green;
     startinternalleavingtlstatus(tuple(l,s)) = green;
     startinternalbarriertlstatus(s) = green;
     startinternalwaterstatus(tuple(l,s)) = unequal;

% Behaviour specification of the lock controller.
proc Controller(BarrierStatus: Position,
                BarrierInEmergency: Bool,
                BarrierLightStatus: StreamSide -> SingleLight,
                GateStatus: LockStreamSideOrientationTriple -> Position,
                PaddleStatus: LockStreamSideOrientationTriple -> Position,
                EnteringLightStatus: LockStreamSideTuple -> DoubleLight,
                LeavingLightStatus: LockStreamSideTuple -> SingleLight,
                WaterLevelStatus: LockStreamSideTuple -> WaterLevel,
                LocksInEmergency: FSet(Lock)) =
% Lock emergency commands.
       sum l:Lock. EmergencyLockCommand(l,activate).
               GateActuator(l, upstream,   east, do_emergencyStop).
               GateActuator(l, upstream,   west, do_emergencyStop).
               GateActuator(l, downstream, east, do_emergencyStop).
               GateActuator(l, downstream, west, do_emergencyStop).
               PaddleActuator(l, upstream,   east, do_emergencyStop).
               PaddleActuator(l, upstream,   west, do_emergencyStop).
               PaddleActuator(l, downstream, east, do_emergencyStop).
               PaddleActuator(l, downstream, west, do_emergencyStop).
               ((EnteringLightStatus(tuple(l,upstream))==redred)
                -> EnteringTrafficLightActuator(l, upstream, east, redred).
                   EnteringTrafficLightActuator(l, upstream, west, redred)
                <> EnteringTrafficLightActuator(l, upstream, east, single_red).
                   EnteringTrafficLightActuator(l, upstream, west, single_red)).
               ((EnteringLightStatus(tuple(l,downstream))==redred)
                -> EnteringTrafficLightActuator(l, downstream, east, redred).
                   EnteringTrafficLightActuator(l, downstream, west, redred)
                <> EnteringTrafficLightActuator(l, downstream, east, single_red).
                   EnteringTrafficLightActuator(l, downstream, west, single_red)).
               LeavingTrafficLightActuator(l, upstream, east, red).
               LeavingTrafficLightActuator(l, upstream, west, red).
               LeavingTrafficLightActuator(l, downstream, east, red).
               LeavingTrafficLightActuator(l, downstream, west, red).
           Controller(LocksInEmergency=LocksInEmergency + {l},
                     EnteringLightStatus=EnteringLightStatus
                        [tuple(l,upstream)->
                           if(EnteringLightStatus(tuple(l,upstream))==redred,
                              redred,single_red)]
                        [tuple(l,downstream)->
                           if(EnteringLightStatus(tuple(l,downstream))==redred,
                              redred,single_red)],
                     LeavingLightStatus=LeavingLightStatus
                        [tuple(l,upstream)->red][tuple(l,downstream)->red]) +
       sum l:Lock. EmergencyLockCommand(l,deactivate).
           Controller(LocksInEmergency = LocksInEmergency - {l}) +
% Gate control.
       sum l:Lock, s:StreamSide. GateCommand(l, s, command_open).
               (GateStatus(triple(l,opposite(s), east)) == closed &&
                GateStatus(triple(l,opposite(s), west)) == closed &&
                PaddleStatus(triple(l,opposite(s), east)) == closed &&
                PaddleStatus(triple(l,opposite(s), west)) == closed &&
                LeavingLightStatus(tuple(l,s)) == red &&
                EnteringLightStatus(tuple(l,s)) != single_green &&
                WaterLevelStatus(tuple(l,s)) == equal &&
                !(l in LocksInEmergency))
                -> sum c_ee: DoubleLightStatus. 
                                 EnteringTrafficLightSensor(l,s,east,c_ee).
                   sum c_ew: DoubleLightStatus. 
                                 EnteringTrafficLightSensor(l,s,west,c_ew).
                   sum c_le: SingleLightStatus. 
                                 LeavingTrafficLightSensor(l,s,east,c_le).
                   sum c_lw: SingleLightStatus. 
                                 LeavingTrafficLightSensor(l,s,west,c_lw).
                   (!(c_ew in { show(single_green), fail_double })  && 
                    !(c_ee in { show(single_green), fail_double }) && 
                    c_lw==show(red) && c_le==show(red))
                   -> GateActuator(l, s, east, do_open).
                      GateActuator(l, s, west, do_open).
                      Controller(GateStatus=GateStatus
                                               [triple(l,s,east)->opening]
                                               [triple(l,s,west)->opening])
                   <> skip.Controller()
                <> skip.Controller()+
       sum l:Lock, s:StreamSide. GateCommand(l, s, command_close).
               (EnteringLightStatus(tuple(l,s)) in { single_red, redred } &&
                LeavingLightStatus(tuple(l,s)) == red &&
                !(l in LocksInEmergency)) 
                -> GateActuator(l, s, east, do_close).
                   GateActuator(l, s, west, do_close).
                   Controller(GateStatus=GateStatus[triple(l,s,east)->closing]
                                                   [triple(l,s,west)->closing])
                <> skip.Controller()+
       sum l:Lock, s:StreamSide. GateCommand(l, s, command_stop).
               GateActuator(l, s, east, do_emergencyStop).
               GateActuator(l, s, west, do_emergencyStop).
               LeavingTrafficLightActuator(l,s,east,red).
               LeavingTrafficLightActuator(l,s,west,red).
               (EnteringLightStatus(tuple(l,s))!=redred)
                -> EnteringTrafficLightActuator(l,s,east,single_red).
                   EnteringTrafficLightActuator(l,s,west,single_red).
                   Controller(EnteringLightStatus=EnteringLightStatus
                                                      [tuple(l,s)->single_red],
                              LeavingLightStatus=LeavingLightStatus
                                                      [tuple(l,s)->red])
                <> EnteringTrafficLightActuator(l,s,east,redred).
                   EnteringTrafficLightActuator(l,s,west,redred).
                   Controller(EnteringLightStatus=EnteringLightStatus
                                                      [tuple(l,s)->redred],
                              LeavingLightStatus=LeavingLightStatus
                                                      [tuple(l,s)->red])+
% Gate sensors.
       sum l:Lock,s:StreamSide,o:Orientation. GateSensor(l,s,o,sense_open).
               (GateStatus(triple(l,s,o)) in { opening, opened }) 
                -> GateActuator(l,s,o, do_endStopOpening).
                   Controller(GateStatus=GateStatus[triple(l,s,o)->opened])
                <> Controller(GateStatus=GateStatus[triple(l,s,o)->closing])+
       sum l:Lock,s:StreamSide,o:Orientation. GateSensor(l,s,o,sense_closed).
               (GateStatus(triple(l,s,o)) in { closing, closed }) 
                -> GateActuator(l,s,o, do_endStopClosing).
                   Controller(GateStatus=GateStatus[triple(l,s,o)->closed])
                <> Controller(GateStatus=GateStatus[triple(l,s,o)->opening])+
       sum l:Lock,s:StreamSide,o:Orientation. 
               (GateSensor(l,s,o,sense_intermediate)+
                GateSensor(l,s,o,fail_position)).
               Controller(GateStatus=
                            if(GateStatus(triple(l,s,o)) in { opening, opened }, 
                               GateStatus[triple(l,s,o)->opening],
                               GateStatus[triple(l,s,o)->closing]))+
% Paddle control.
       sum l:Lock, s:StreamSide. PaddleCommand(l, s, command_open).
               (GateStatus(triple(l,opposite(s), east)) == closed &&
                GateStatus(triple(l,opposite(s), west)) == closed &&
                PaddleStatus(triple(l,opposite(s), east)) == closed &&
                PaddleStatus(triple(l,opposite(s), west)) == closed &&
                !(l in LocksInEmergency)) 
                -> PaddleActuator(l, s, east, do_open).
                   PaddleActuator(l, s, west, do_open).
                   Controller(PaddleStatus=
                              PaddleStatus[triple(l,s,east)->opening]
                                          [triple(l,s,west)->opening])
                <> skip.Controller()+
       sum l:Lock, s:StreamSide. PaddleCommand(l, s, command_close).
               (!(l in LocksInEmergency)) 
                -> PaddleActuator(l, s, east, do_close).
                   PaddleActuator(l, s, west, do_close).
                   Controller(PaddleStatus=
                              PaddleStatus[triple(l,s,east)->closing]
                                          [triple(l,s,west)->closing])
                <> skip.Controller()+
       sum l:Lock, s:StreamSide. PaddleCommand(l, s, command_stop).
               PaddleActuator(l, s, east, do_emergencyStop).
               PaddleActuator(l, s, west, do_emergencyStop).
               Controller()+
% Paddle sensors.
       sum l:Lock,s:StreamSide,o:Orientation. PaddleSensor(l,s,o,sense_open).
               (PaddleStatus(triple(l,s,o)) in { opening, opened }) 
                -> PaddleActuator(l,s,o, do_endStopOpening).
                   Controller(PaddleStatus=PaddleStatus[triple(l,s,o)->opened])
                <> Controller(PaddleStatus=PaddleStatus[triple(l,s,o)->closing])+
       sum l:Lock,s:StreamSide,o:Orientation. PaddleSensor(l,s,o,sense_closed).
               (PaddleStatus(triple(l,s,o)) in { closing, closed }) 
                -> PaddleActuator(l,s,o, do_endStopClosing).
                   Controller(PaddleStatus=PaddleStatus[triple(l,s,o)->closed])
                <> Controller(PaddleStatus=PaddleStatus[triple(l,s,o)->opening])+
       sum l:Lock,s:StreamSide,o:Orientation. 
                (PaddleSensor(l,s,o,sense_intermediate)+
                 PaddleSensor(l,s,o,fail_position)).  
                Controller(PaddleStatus=
                    if(PaddleStatus(triple(l,s,o)) in { opening, opened }, 
                        PaddleStatus[triple(l,s,o)->opening],
                        PaddleStatus[triple(l,s,o)->closing]))+
% Water sensors.
       sum l:Lock,s:StreamSide. WaterSensor(l,s,equal).
                Controller(WaterLevelStatus=WaterLevelStatus[tuple(l,s)->equal])+
       sum l:Lock,s:StreamSide. 
                (WaterSensor(l,s,unequal)+WaterSensor(l,s,fail_water_sensor)).
                Controller(WaterLevelStatus=WaterLevelStatus[tuple(l,s)->unequal])+
% Lock entering traffic light control.
       sum l:Lock, s:StreamSide. EnteringTrafficLightCommand(l,s,redred).
                EnteringTrafficLightActuator(l,s,east,redred).
                EnteringTrafficLightActuator(l,s,west,redred).
                Controller(EnteringLightStatus=
                           EnteringLightStatus[tuple(l,s)->redred])+
       sum l:Lock, s:StreamSide. EnteringTrafficLightCommand(l, s, single_red).
                EnteringTrafficLightActuator(l,s,east,single_red).
                EnteringTrafficLightActuator(l,s,west,single_red).
                Controller(EnteringLightStatus=
                           EnteringLightStatus[tuple(l,s)->single_red])+
       sum l:Lock, s:StreamSide. EnteringTrafficLightCommand(l, s, redgreen).
               ( LeavingLightStatus(tuple(l,s)) == red) 
                -> EnteringTrafficLightActuator(l,s,east,redgreen).
                   EnteringTrafficLightActuator(l,s,west,redgreen).
                   Controller(EnteringLightStatus=
                              EnteringLightStatus[tuple(l,s)->redgreen])
                <> skip.Controller()+
       sum l:Lock, s:StreamSide. EnteringTrafficLightCommand(l, s, single_green).
               (LeavingLightStatus(tuple(l,s)) == red &&
                GateStatus(triple(l,s,east)) == opened &&
                GateStatus(triple(l,s,west)) == opened) 
                -> sum c_east: SingleLightStatus. 
                                     LeavingTrafficLightSensor(l,s,east,c_east).
                   sum c_west: SingleLightStatus. 
                                     LeavingTrafficLightSensor(l,s,west,c_west).
                   (c_east==show(red) && c_west==show(red))
                   -> EnteringTrafficLightActuator(l,s,east,single_green).
                      EnteringTrafficLightActuator(l,s,west,single_green).
                      Controller(EnteringLightStatus=
                                 EnteringLightStatus[tuple(l,s)->single_green])
                   <> skip.Controller()
                <> skip.Controller()+
% Lock leaving traffic light control.
       sum l:Lock, s:StreamSide. LeavingTrafficLightCommand(l, s, red).
                LeavingTrafficLightActuator(l,s,east,red).
                LeavingTrafficLightActuator(l,s,west,red).
                Controller(LeavingLightStatus=
                           LeavingLightStatus[tuple(l,s)->red])+
       sum l:Lock, s:StreamSide. LeavingTrafficLightCommand(l, s, green).
               (EnteringLightStatus(tuple(l,s)) in { single_red, redred } &&
                GateStatus(triple(l,s,east)) == opened &&
                GateStatus(triple(l,s,west)) == opened) 
                -> sum c_east: DoubleLightStatus. 
                                     EnteringTrafficLightSensor(l,s,east,c_east).
                   sum c_west: DoubleLightStatus. 
                                     EnteringTrafficLightSensor(l,s,west,c_west).
                   (c_east in {show(single_red), show(redred)} && 
                    c_west in {show(single_red), show(redred)})
                    -> LeavingTrafficLightActuator(l,s,east,green).
                       LeavingTrafficLightActuator(l,s,west,green).
                       Controller(LeavingLightStatus=
                                  LeavingLightStatus[tuple(l,s)->green])
                    <> skip.Controller()
                <> skip.Controller()+
% Barrier emergency commands.
       EmergencyBarrierCommand(activate).
               BarrierActuator(do_emergencyStop).
               BarrierTrafficLightActuator(upstream,east,red).
               BarrierTrafficLightActuator(upstream,west,red).
               BarrierTrafficLightActuator(downstream,east,red).
               BarrierTrafficLightActuator(downstream,west,red).
               Controller(BarrierInEmergency=true,
                          BarrierLightStatus=
                              BarrierLightStatus[upstream->red][downstream->red]) +
       EmergencyBarrierCommand(deactivate).
               Controller(BarrierInEmergency=false) +
% Barrier control
       BarrierCommand(command_open).
               (BarrierLightStatus(upstream) == red &&
                BarrierLightStatus(downstream) == red &&
                !BarrierInEmergency)
                -> sum c_ue:SingleLightStatus. 
                                  BarrierTrafficLightSensor(upstream,east,c_ue). 
                   sum c_uw:SingleLightStatus. 
                                  BarrierTrafficLightSensor(upstream,west,c_uw).
                   sum c_de:SingleLightStatus. 
                                  BarrierTrafficLightSensor(downstream,east,c_de). 
                   sum c_dw:SingleLightStatus. 
                                  BarrierTrafficLightSensor(downstream,west,c_dw).
                   (c_ue == show(red) && c_uw == show(red) && 
                    c_de == show(red) && c_dw == show(red))
                    -> BarrierActuator(do_open).
                       Controller(BarrierStatus=opening)
                    <> skip.Controller()
                <> skip.Controller() +
       BarrierCommand(command_close).
               (BarrierLightStatus(upstream) == red &&
                BarrierLightStatus(downstream) == red &&
                !BarrierInEmergency)
                -> BarrierActuator(do_close).
                   Controller(BarrierStatus=closing)
                <> skip.Controller() +
       BarrierCommand(command_stop).
               BarrierTrafficLightActuator(upstream,east,red).
               BarrierTrafficLightActuator(upstream,west,red).
               BarrierTrafficLightActuator(downstream,east,red).
               BarrierTrafficLightActuator(downstream,west,red).
               BarrierActuator(do_emergencyStop). 
               Controller(BarrierLightStatus=
                              BarrierLightStatus[upstream->red][downstream->red]) +
% Barrier traffic light.
       sum s:StreamSide. BarrierTrafficLightCommand(s, red).
                BarrierTrafficLightActuator(s,east,red).
                BarrierTrafficLightActuator(s,west,red).
                Controller(BarrierLightStatus=BarrierLightStatus[s->red])+
       sum s:StreamSide. BarrierTrafficLightCommand(s, green).
               (BarrierStatus == opened)
                -> BarrierTrafficLightActuator(s,east,green).
                   BarrierTrafficLightActuator(s,west,green).
                   Controller(BarrierLightStatus=BarrierLightStatus[s->green])
                <> skip.Controller()+
% Barrier sensor.
       BarrierSensor(sense_open).
               (BarrierStatus in { opening, opened })
                -> BarrierActuator(do_endStopOpening).
                   Controller(BarrierStatus=opened)
                <> Controller(BarrierStatus=closing)+
       BarrierSensor(sense_closed).
               (BarrierStatus in { closing, closed })
                -> BarrierActuator(do_endStopClosing).
                   Controller(BarrierStatus=closed)
                <> Controller(BarrierStatus=opening)+
       (BarrierSensor(sense_intermediate)+BarrierSensor(fail_position)).
                   Controller(BarrierStatus=
                       if(BarrierStatus in { opening, opened }, opening, closing));

% Initial state of the controller.
init Controller(opening, 
                false,
                startinternalbarriertlstatus,
                startinternallockstatus, 
                startinternalpaddlestatus,
                startinternalenteringtlstatus, 
                startinternalleavingtlstatus, 
                startinternalwaterstatus,
                {});
\end{lstlisting}
\section{The requirements as modal formulas}
\label{app:formulas}
In this appendix we provide the translations of the requirements as listed in Section~\ref{sec:requirements} into modal formulas. 
We only compactly explain the structure of the modal formulas. 
A full explanation of modal formulas in general can be found in for instance~\cite{DBLP:books/mit/GrooteM2014}.
\subsection{Safety requirements}
All the safety requirements have the following shape.
\begin{lstlisting}[language=mCRL2]
          [true*.a.!b*.c]false && [!b*.c]false
\end{lstlisting}
omitting details such as quantification. The first part before the conjunction
says that
it is not possible, expressed by the false at the end, to execute a trace
starting with an arbitrary sequence of actions (\formula{true*}), followed
by an action \formula{a}, followed by a sequence of actions not containing
\formula{b} (\formula{!b*}), followed by an action \formula{c}. In words
this reads ``whenever an action \formula{a} happens, an action \formula{c},
can only follow if it is preceded by an action \formula{b}''.

The second part expresses that an action \formula{c} cannot be performed
if it is not preceded by a sequence of actions containing action \formula{b}.
In other words, action \formula{c} has to be preceded by an action \formula{b}. 

At certain places below we write \formula{a||b}. This means either action \formula{a} or action \formula{b}. We also use 
\formula{val(cond) => ...} which 
represents logical implication. If the condition \formula{cond} is valid, the
formula at the dots must hold. 

\subsubsection{Requirement~\ref{safreq1}: Opposing paddles cannot be both open simultaneously}
This property is concretely translated to a formula saying that it is not possible to instruct a paddle to open,
if no \formula{sense\_closed} of both paddles in the opposing gates have been observed after
they were sensed to be not closed, sensing failed, or after they were instructed to open. Initially, a paddle can also only
open, if the opposing paddles have been measured to be closed. 
\begin{lstlisting}[language=mCRL2]
% If the paddles are not closed at one stream side of a sluice, the paddles 
% at the other stream side cannot be opened. 

(forall l:Lock, s:StreamSide, o:Orientation.
  [true*.
   (PaddleSensor(l,s,o,sense_intermediate)||PaddleSensor(l,s,o,sense_open)||
        PaddleSensor(l,s,o,fail_position)||PaddleActuator(l,s,o,do_open)).
   !(PaddleSensor(l,s,o,sense_closed))*.
   exists o':Orientation.PaddleActuator(l,opposite(s),o',do_open)]false)
&&
forall l:Lock, s:StreamSide, o: Orientation.
  [!(PaddleSensor(l,s,o,sense_closed))*.
   exists o':Orientation.PaddleActuator(l,opposite(s),o',do_open)]false
\end{lstlisting}
\subsubsection{Requirement~\ref{safreq2}: Paddles cannot open with an opposing gate open}
The modal formula representing this property says that a sequences of actions where a gate in a lock is 
measured to not be closed, the sensor failed, or is actively opened, not followed by a measurement that it is closed,
cannot be followed by opening the paddle in an opposing gate.
\begin{lstlisting}[language=mCRL2]
% If the gates are not closed at one stream side of a sluice, the paddles 
% at the other stream side cannot be opened.

forall l:Lock, s:StreamSide, o :Orientation.
  [true*.
   (GateSensor(l,s,o,sense_intermediate)||GateSensor(l,s,o,sense_open)||
         GateSensor(l,s,o,fail_position)||GateActuator(l,s,o,do_open)).
   !(GateSensor(l,s,o,sense_closed))*.
   exists o':Orientation.PaddleActuator(l,opposite(s),o',do_open)]false
&&
forall l:Lock, s:StreamSide, o: Orientation.
  [!(GateSensor(l,s,o,sense_closed))*.
   exists o':Orientation.PaddleActuator(l,opposite(s),o',do_open)]false
\end{lstlisting}
\subsubsection{Requirement~\ref{safreq3}: Gates cannot open with an opposing paddle open}
The formula has the same shape as the previous formulas and says that it is not possible that
after the paddles have been measured not to be closed, a measurement failed, 
or if the paddles have been instructed to open, the opposing gates cannot be instructed to open
unless the paddles have been measured to be closed in the meantime. Moreover, the gates in a lock
cannot be opened initially, without having measured that the opposing paddles are closed. 
\begin{lstlisting}[language=mCRL2]
% If the paddles are not closed at one stream side of a sluice, the gates at 
% the other stream side cannot be opened.

forall l:Lock, s:StreamSide, o:Orientation.
  [true*.
   (PaddleSensor(l,s,o,sense_open)||PaddleSensor(l,s,o,sense_intermediate)||
         PaddleSensor(l,s,o,fail_position)||PaddleActuator(l,s,o,do_open)).
   !(PaddleSensor(l,s,o,sense_closed))*.
   exists o':Orientation.GateActuator(l,opposite(s),o',do_open)]false
&&
forall l:Lock, s:StreamSide, o:Orientation.
  [!(PaddleSensor(l,s,o,sense_closed))*.
   exists o':Orientation.GateActuator(l,opposite(s),o',do_open)]false
\end{lstlisting}
\subsubsection{Requirement~\ref{safreq4}: Gates cannot open with an opposing gate open}
The translation to the modal formula says that if a gate in a lock is not measured to be closed, 
such a measurement failed, or the gate is instructed to be open, and the gate is subsequently not measured
to be closed, then an opposite gate cannot be opened. Also, initially a gate cannot be instructed to open,
if an opposing gate is not known to be closed. 
\begin{lstlisting}[language=mCRL2]
% If the gates are not closed at one stream side of a sluice, the gates 
% at the opposing stream side cannot be opened.

forall l:Lock, s:StreamSide, o:Orientation.
  [true*.
   (GateSensor(l,s,o,sense_intermediate)||GateSensor(l,s,o,sense_open)||
      GateSensor(l,s,o,fail_position)||GateActuator(l,s,o,do_open)).
   !(GateSensor(l,s,o,sense_closed))*.
   exists o':Orientation.GateActuator(l,opposite(s),o',do_open)]false
&&
forall l:Lock, s:StreamSide, o:Orientation.
  [!(GateSensor(l,s,o,sense_closed))*.
   exists o':Orientation.GateActuator(l,opposite(s),o',do_open)]false
\end{lstlisting}
\subsubsection{Requirement~\ref{safreq5}: Gates can only open if the waterlevel is equal}
This is formalised by the modal formula that says that if a gate in a lock is instructed to open, then
a measurement that the water level was equal must have happened
after a measurement failed, or it indicated unequal water levels around that gate. 
Initially, a gate can only be instructed to open, if it has been measured at least once 
that the water level around the gate is equal. 
\begin{lstlisting}[language=mCRL2]
% If the water level at a stream side of a sluice is not level, 
% then both the east and west doors may not be opened.

forall l:Lock, s:StreamSide.
  [true*.
   (WaterSensor(l,s,unequal)||WaterSensor(l,s,fail_water_sensor)).
   !(WaterSensor(l,s,equal))*.
   exists o:Orientation.GateActuator(l,s,o,do_open)]false
&&
forall l:Lock, s:StreamSide.
  [!(WaterSensor(l,s,equal))*.
   exists o: Orientation.GateActuator(l,s,o,do_open)]false
\end{lstlisting}
\subsubsection{Requirement~\ref{safreq6}: Traffic lights at entering and leaving side I}
This is formalised by indicating that the entering lights of a lock can only be set to single green if
after the corresponding leaving lights have been measured to show green or fail, or have been instructed to change,
the leaving light must have been measured to be red. Also, initially, before setting the entering light to
single green, the corresponding leaving lights must have been measured to show red. 
\begin{lstlisting}[language=mCRL2]
% If the traffic lights at the entering side of a sluice are set to single green, 
% they must be red at the leaving side. 

forall l:Lock, s:StreamSide, o:Orientation.
  [true*.
   (LeavingTrafficLightSensor(l,s,o,show(green))||
         LeavingTrafficLightSensor(l,s,o,fail_single)||
         exists c:SingleLight.LeavingTrafficLightActuator(l,s,o,c)).
   !(LeavingTrafficLightSensor(l,s,o,show(red)))*.
   exists o':Orientation.EnteringTrafficLightActuator(l,s,o',single_green)]false
&&
forall l:Lock, s:StreamSide, o: Orientation.
  [!(LeavingTrafficLightSensor(l,s,o,show(red)))*.
   exists o':Orientation.EnteringTrafficLightActuator(l,s,o',single_green)]false
\end{lstlisting}
\subsubsection{Requirement~\ref{safreq13}: Traffic lights at entering and leaving side II}
This is translated to a formula that says that if the entering traffic lights are set to single green,
then the gates of this side of the lock must have been measured to be open, after the last measurement
that shows that the gates were not open, or this last measurement failed, or the gates have been instructed to
move. Initially, the lights can only be set to single green, if it has been measured that the gates were open.

\begin{lstlisting}[language=mCRL2]
% If the traffic lights at the leaving side of a sluice are set to green,
% they must be single red or red-red at the entering side. 

forall l:Lock, s:StreamSide, o:Orientation, a:DoubleLight.
 (val(a in { single_green, redgreen})=>
  [true*.
   (EnteringTrafficLightSensor(l,s,o,show(a))||
        EnteringTrafficLightSensor(l,s,o,fail_double)||
        exists c:DoubleLight.EnteringTrafficLightActuator(l,s,o,c)).
   !(exists a':DoubleLight.val(a' in {single_red, redred}) && 
                   EnteringTrafficLightSensor(l,s,o,show(a')))*.
   exists o': Orientation.LeavingTrafficLightActuator(l,s,o',green)]false)
&&
forall l:Lock, s:StreamSide.
  [!(exists o: Orientation, a:DoubleLight.val(a in {single_red, redred}) && 
        EnteringTrafficLightSensor(l,s,o,show(a)))*.
   exists o': Orientation.LeavingTrafficLightActuator(l,s,o',green)]false
\end{lstlisting}
\subsubsection{Requirement~\ref{safreq7}: Lights cannot be set to green if lock not open I}
This formula expresses that the entering traffic lights cannot be set to green if the gates
of the lock have not been measured to be open. 
\begin{lstlisting}[language=mCRL2]
% If a gate is not open, the entering traffic lights cannot be changed to green.

forall l:Lock, s:StreamSide, o:Orientation.
  [true*. 
   (GateSensor(l,s,o,sense_closed)||GateSensor(l,s,o,sense_intermediate)||
    GateSensor(l,s,o,fail_position)||GateActuator(l,s,o,do_close)).
   !(GateSensor(l,s,o,sense_open))*.
   exists o':Orientation.EnteringTrafficLightActuator(l,s,o',single_green)]false
&&
forall l:Lock, s:StreamSide, o:Orientation.
  [!(GateSensor(l,s,o,sense_open))*.
   exists o':Orientation.EnteringTrafficLightActuator(l,s,o',single_green)]false
\end{lstlisting}
\subsubsection{Requirement~\ref{safreq14}: Lights cannot be set to green if lock not open II}
This formula has the same structure as the previous formula, except that it now regards the leaving lights. 
\begin{lstlisting}[language=mCRL2]
% If a gate is not open, the leaving traffic lights cannot be set to green.

forall l:Lock, s:StreamSide, o:Orientation.
  [true*. 
   (GateSensor(l,s,o,sense_closed)||GateSensor(l,s,o,sense_intermediate)||
        GateSensor(l,s,o,fail_position)||GateActuator(l,s,o,do_close)).
   !(GateSensor(l,s,o,sense_open))*.
   exists o':Orientation.LeavingTrafficLightActuator(l,s,o',green)]false
&&
forall l:Lock, s:StreamSide, o:Orientation.
  [!(GateSensor(l,s,o,sense_open))*.
   exists o':Orientation.LeavingTrafficLightActuator(l,s,o',green)]false 
\end{lstlisting}
\subsubsection{Requirement~\ref{safreq8}: Gates cannot be closed if the lights are not set to red I}
The translation expresses that the gates can only be instructed to close if
after changing the entering lights to another colour than red or red-red, they have been set to single red or red-red again.
Initially, the gates can only close if the entering traffic lights have been set at least once to single red or red-red.

\begin{lstlisting}[language=mCRL2]
% If the entering traffic lights are not set to single red or redred, 
% the doors cannot be instructed to close. 

forall l:Lock, s:StreamSide, o:Orientation, a:DoubleLight.
 (val(a in {single_green, redgreen})=>
  [true*.
   (EnteringTrafficLightActuator(l,s,o,a)).
   !(exists a':DoubleLight.(val(a' in {single_red, redred}) && 
         EnteringTrafficLightActuator(l,s,o,a')))*.
   exists o':Orientation.GateActuator(l,s,o',do_close)]false)
&&
forall l:Lock, s:StreamSide, o: Orientation.
  [!(exists a:DoubleLight.(val(a in {single_red, redred}) && 
         EnteringTrafficLightActuator(l,s,o,a)))*. 
   exists o':Orientation.GateActuator(l,s,o',do_close)]false
\end{lstlisting}
\subsubsection{Requirement~\ref{safreq9}: Gates cannot be closed if the lights are not set to red II}
This requirement is very similar to the previous requirement except that it is now about the leaving lights. 
\begin{lstlisting}[language=mCRL2]
% If the leaving traffic lights are not set red, the gates cannot be 
% instructed to close. 

forall l:Lock, s:StreamSide, o:Orientation. 
  [true*.
   LeavingTrafficLightActuator(l,s,o,green).
   !(LeavingTrafficLightActuator(l,s,o,red))*.
   exists o':Orientation.GateActuator(l,s,o',do_close)]false &&

forall l:Lock, s:StreamSide, o:Orientation.
  [!(LeavingTrafficLightActuator(l,s,o,red))*. 
   exists o':Orientation.GateActuator(l,s,o',do_close)]false
\end{lstlisting}
\subsubsection{Requirement~\ref{safreq10}: Gates and paddles cannot move in emergency mode}
This requirement is translated into a formula that says that if the emergency mode for a lock has been
activated, and subsequently not deactivated, then it is not possible to open or close a gate or paddle. 
\begin{lstlisting}[language=mCRL2]
% In emergency mode the gates and paddles cannot be instructed to open or close.

forall l:Lock.
  [true*.
   (EmergencyLockCommand(l,activate)).
   !(EmergencyLockCommand(l,deactivate))*.
   exists s:StreamSide, o:Orientation.
       (GateActuator(l,s,o,do_close)|| GateActuator(l,s,o,do_open)||
        PaddleActuator(l,s,o,do_close)||PaddleActuator(l,s,o,do_open))]false
\end{lstlisting}
\subsubsection{Requirement~\ref{safreq23}\label{transsafreq23}: End stop opening gate only if open}
This formula says that it is only possible to send a \formula{do\_endStopOpening} to a gate if
it is measured to be open after measuring that it was not open or a measurement has failed. Initially,
an \formula{do\_endStopOpening} can be sent if the gate has been measured to be open at least once. 
\begin{lstlisting}[language=mCRL2]
% If a gate is instructed to do an endStopOpening, it is known to be open. 

forall l:Lock, s:StreamSide, o:Orientation.
  [true*.
   (GateSensor(l,s,o,sense_intermediate)||GateSensor(l,s,o,sense_closed)||
        GateSensor(l,s,o,fail_position)).
   !GateSensor(l,s,o,sense_open)*.
   GateActuator(l,s,o,do_endStopOpening)]false
&&
forall l:Lock, s:StreamSide, o:Orientation.
  [!GateSensor(l,s,o,sense_open)*.
   GateActuator(l,s,o,do_endStopOpening)]false
\end{lstlisting}
\subsubsection{Requirement~\ref{safreq24}: End stop closing gate only if closed}
This requirement is similar to the previous one except that it is now about sending the close command.
\begin{lstlisting}[language=mCRL2]
% If a gate is instructed to do an endStopClosing, it is known to be closed.

forall l:Lock, s:StreamSide, o:Orientation.
  [true*.
   (GateSensor(l,s,o,sense_intermediate)||GateSensor(l,s,o,sense_open)||
       GateSensor(l,s,o,fail_position)).
   !GateSensor(l,s,o,sense_closed)*.
   GateActuator(l,s,o,do_endStopClosing)]false
&&
forall l:Lock, s:StreamSide, o:Orientation.
  [!GateSensor(l,s,o,sense_closed)*.
   GateActuator(l,s,o,do_endStopClosing)]false
\end{lstlisting}
\subsubsection{Requirement~\ref{safreq27}: End stop opening paddle only if open}
This is similar to Requirement~\ref{transsafreq23}.
\begin{lstlisting}[language=mCRL2]
% If a paddle is instructed to do an endStopOpening, it is known to be open. 

forall l:Lock, s:StreamSide, o:Orientation.
  [true*.
   (PaddleSensor(l,s,o,sense_closed)||PaddleSensor(l,s,o,sense_intermediate)||
       PaddleSensor(l,s,o,fail_position)).
   !PaddleSensor(l,s,o,sense_open)*.
   PaddleActuator(l,s,o,do_endStopOpening)]false
&&
forall l:Lock, s:StreamSide, o:Orientation.
  [!PaddleSensor(l,s,o,sense_open)*.
   PaddleActuator(l,s,o,do_endStopOpening)]false
\end{lstlisting}
\subsubsection{Requirement~\ref{safreq28}: End stop closing gate only if closed}
See Requirement~\ref{transsafreq23}.
\begin{lstlisting}[language=mCRL2]
% If a paddle is instructed to do an endStopClosing, it is known to be closed.

forall l:Lock, s:StreamSide, o:Orientation.
  [true*.
   (PaddleSensor(l,s,o,sense_intermediate)||PaddleSensor(l,s,o,sense_open)||
       PaddleSensor(l,s,o,fail_position)).
   !PaddleSensor(l,s,o,sense_closed)*.
   PaddleActuator(l,s,o,do_endStopClosing)]false
&&
forall l:Lock, s:StreamSide, o:Orientation.
  [!PaddleSensor(l,s,o,sense_closed)*.
   PaddleActuator(l,s,o,do_endStopClosing)]false
\end{lstlisting}
\subsubsection{Requirement~\ref{safreq29}: Barrier only closes when lights are red}
This modal formulas says that, both initially, and after setting a traffic light of the barrier to green, it must
first have been set to red, before the barrier actuator can be instructed to open. 
\begin{lstlisting}[language=mCRL2]
% If the barrier actuator is instructed to close, then both the upstream and 
% downstream traffic lights are set to red.

forall s:StreamSide, o:Orientation.
  [true*.
   BarrierTrafficLightActuator(s,o,green).
   !BarrierTrafficLightActuator(s,o,red)*.
   BarrierActuator(do_close)]false
&&
forall s:StreamSide, o:Orientation.
  [!BarrierTrafficLightActuator(s,o,red)*.
   BarrierActuator(do_close)]false
\end{lstlisting}
\subsubsection{Requirement~\ref{safreq30}: Barrier lights only become green when the barrier is open}
This is translated to a modal formula by requiring that if the barrier is not sensed to be open, i.e., is closed, in an intermediate
position, the measurement fails, or the system is in the initial state, then the barrier must be measured to be open,
before the actuators of the lights can be instructed to go to green. 
\begin{lstlisting}[language=mCRL2]
% If the barrier traffic lights are set to green, the barrier is open.

  [true*.
   (BarrierSensor(sense_closed)||BarrierSensor(sense_intermediate)||
       BarrierSensor(fail_position)).
   !BarrierSensor(sense_open)*.
   exists s:StreamSide, o:Orientation.BarrierTrafficLightActuator(s,o,green)]false
&&
  [!BarrierSensor(sense_open)*.
   exists s:StreamSide, o:Orientation.BarrierTrafficLightActuator(s,o,green)]false
\end{lstlisting}
\subsubsection{Requirement~\ref{safreq40}: The barrier cannot move in emergency mode}
This follows the standard pattern which says that whenever an emergency of the barrier is activated,
this cannot be followed by an open or close command, unless a deactivate from emergency mode came first. 
\begin{lstlisting}[language=mCRL2]
% In emergency mode the barrier cannot be instructed to open or close.

[true*.
  EmergencyBarrierCommand(activate).
  !(EmergencyBarrierCommand(deactivate))*.
  (BarrierActuator(do_close)||BarrierActuator(do_open))]false
\end{lstlisting}

\subsubsection{Requirement~\ref{safreq35}: End stop opening barrier only if the barrier is open}
This requirement is translated in a similar way as Requirement~\ref{transsafreq23}.
\begin{lstlisting}[language=mCRL2]
% If the barrier is instructed to do an endStopOpening, it is open.

forall u:SensorPosition.(val(u!=sense_open) =>
  [true*.
   BarrierSensor(u).
   !BarrierSensor(sense_open)*.
   BarrierActuator(do_endStopOpening)]false)
&&
  [!BarrierSensor(sense_open)*.
   BarrierActuator(do_endStopOpening)]false
\end{lstlisting}
\subsubsection{Requirement~\ref{safreq36}: End stop closing barrier only if the barrier is closed}
This requirement is translated similarly to Requirement~\ref{transsafreq23}.
\begin{lstlisting}[language=mCRL2]
% If the barrier is instructed to do an endStopClosing, it is closed.

forall u:SensorPosition.(val(u!=sense_closed) =>
  [true*.
   BarrierSensor(u).
   !BarrierSensor(sense_closed)*.
   BarrierActuator(do_endStopClosing)]false)
 &&
  [!BarrierSensor(sense_closed)*.
   BarrierActuator(do_endStopClosing)]false
\end{lstlisting}
\subsection{Causality requirements}
All the causality requirements are translated to modal formulas in the same way, so we do not explain each 
translation separately. Basically, the translation follows the same structure as the safety requirements.
says that As an example consider Requirement~\ref{transcausreq15}. It says that a gate cannot open
without the operator having given the open command. 
The translation says that after each instruction sent to the gate actuator,
a \formula{do\_open} command can not be sent to the actuator unless a \formula{command\_open}
has been received from the console, after any previous command that has been sent to the gate actuator. 
Moreover, in the initial state, opening a gate actuator is only possible if at least one 
open command from the console has been received. 
\subsubsection{Requirement~\ref{causreq11}\label{transcausreq11}: Emergency stop of a gate}
\begin{lstlisting}[language=mCRL2]
% If an emergency stop of a gate in a lock takes place, this has to be preceded 
% by either an emergency stop of that lock, or the instruction to stop that
% gate, which happened after the last instruction to that gate.

forall l:Lock, s:StreamSide, o:Orientation, c:ActuatorCommand. 
  [true*.
   GateActuator(l,s,o,c).
   !(EmergencyLockCommand(l,activate)||GateCommand(l,s,command_stop))*.
   GateActuator(l,s,o,do_emergencyStop)]false
&&
forall l:Lock, s:StreamSide.
  [!(EmergencyLockCommand(l,activate)||GateCommand(l,s,command_stop))*.
   exists o: Orientation.GateActuator(l,s,o,do_emergencyStop)]false
\end{lstlisting}
\subsubsection{Requirement~\ref{causreq12}: Emergency stop of a paddle}
\begin{lstlisting}[language=mCRL2]
% An emergency stop of a paddle in a lock has to be preceded 
% by either an emergency stop of that lock, or the instruction to stop that
% paddle, which happened after the last instruction to that paddle.

forall l:Lock, s:StreamSide, o:Orientation, c:ActuatorCommand. 
  [true*.
    (PaddleActuator(l,s,o,c)).
    !(EmergencyLockCommand(l,activate)||PaddleCommand(l,s,command_stop))*.
    PaddleActuator(l,s,o,do_emergencyStop)]false
&&
forall l:Lock, s:StreamSide.
  [!(EmergencyLockCommand(l,activate)||PaddleCommand(l,s,command_stop))*.
   exists o:Orientation.PaddleActuator(l,s,o,do_emergencyStop)]false
\end{lstlisting}
\subsubsection{Requirement~\ref{causreq15}\label{transcausreq15}: Opening a gate}
\begin{lstlisting}[language=mCRL2]
% If a gate of a lock is opened, this has to be preceded by an instruction to open 
% that gate, and this instruction must have happened after the last command sent 
% to the gate.

forall l:Lock, s:StreamSide, o:Orientation, c:ActuatorCommand.
  [true*.
   (GateActuator(l,s,o,c)).
   !(GateCommand(l,s,command_open))*.
   GateActuator(l,s,o,do_open)]false
&&
forall l:Lock, s:StreamSide, o: Orientation.
  [!(GateCommand(l,s,command_open))*.
   GateActuator(l,s,o,do_open)]false
\end{lstlisting}
\subsubsection{Requirement~\ref{causreq16}: Closing a gate}
\begin{lstlisting}[language=mCRL2]
% If a gate of a lock is closed, this has to be preceded by an instruction to close 
% that gate, and this instruction must have happened after the last command sent to 
% the gate.

forall l:Lock, s:StreamSide, o:Orientation, c:ActuatorCommand.
  [true*.
   (GateActuator(l,s,o,c)).
   !(GateCommand(l,s,command_close))*.
   GateActuator(l,s,o,do_close)]false
&&
forall l:Lock, s:StreamSide, o: Orientation.
  [!(GateCommand(l,s,command_close))*.
   GateActuator(l,s,o,do_close)]false
\end{lstlisting}
\subsubsection{Requirement~\ref{causreq17}: Opening a paddle}
\begin{lstlisting}[language=mCRL2]
% If a paddle of a lock is opened, this has to be preceded by an instruction to 
% open that paddle, and this instruction must have happened after the last 
% command sent to the paddle.

forall l:Lock, s:StreamSide, o:Orientation, c:ActuatorCommand.
  [true*.
   (PaddleActuator(l,s,o,c)).
   !(PaddleCommand(l,s,command_open))*.
   PaddleActuator(l,s,o,do_open)]false
&&
forall l:Lock, s:StreamSide, o: Orientation.
  [!(PaddleCommand(l,s,command_open))*.
   PaddleActuator(l,s,o,do_open)]false
\end{lstlisting}
\subsubsection{Requirement~\ref{causreq18}: Closing a paddle}
\begin{lstlisting}[language=mCRL2]
% If a paddle of a lock is closed, this has to be preceded by an instruction to 
% close that paddle, and this instruction must have happened after the last 
% command sent to the paddle.

forall l:Lock, s:StreamSide, o:Orientation, c:ActuatorCommand.
  [true*.
   PaddleActuator(l,s,o,c).
   !(PaddleCommand(l,s,command_close))*.
   PaddleActuator(l,s,o,do_close)]false
&&
forall l:Lock, s:StreamSide, o: Orientation.
  [!(PaddleCommand(l,s,command_close))*.
   PaddleActuator(l,s,o,do_close)]false
\end{lstlisting}
\subsubsection{Requirement~\ref{causreq19}: Setting the entering lights in a lock}
\begin{lstlisting}[language=mCRL2]
% If the traffic lights at the entering side of a sluice are set to some color a', 
% an instruction to do so must have been given since the last instruction to the 
% traffic lights.  In case the lights go to single_red, this can also be caused by 
% activating an emergency or stopping a gate. 

(forall l:Lock, s:StreamSide, o:Orientation, a,a':DoubleLight.
 (val(a' in { single_green, redgreen }) =>
  [true*.
   EnteringTrafficLightActuator(l,s,o,a).
   !EnteringTrafficLightCommand(l,s,a')*.
   EnteringTrafficLightActuator(l,s,o,a')]false))
&&
(forall l:Lock, s:StreamSide, o:Orientation, a,a':DoubleLight.
 (val(a' in { single_red, redred }) =>
  [true*.
   (EnteringTrafficLightActuator(l,s,o,a)).
   !(EnteringTrafficLightCommand(l,s,single_red)||
       EmergencyLockCommand(l,activate)||
       GateCommand(l, s, command_stop))*.
   EnteringTrafficLightActuator(l,s,o,single_red)]false))
&&
(forall l:Lock, s:StreamSide, a:DoubleLight.
 (val(a in { single_green, redgreen, redred }) =>
  [!EnteringTrafficLightCommand(l,s,a)*.
   exists o:Orientation.EnteringTrafficLightActuator(l,s,o,a)]false))
&&
(forall l:Lock, s:StreamSide, a:DoubleLight.
 (val(a in { single_red, redred }) =>
  [!(EnteringTrafficLightCommand(l,s,a)||
       EmergencyLockCommand(l,activate)||
       GateCommand(l, s, command_stop))*.
   exists o:Orientation.EnteringTrafficLightActuator(l,s,o,a)]false))
\end{lstlisting}
\subsubsection{Requirement~\ref{causreq20}: Setting the leaving lights in a lock}
\begin{lstlisting}[language=mCRL2]
% If the traffic lights at the leaving side of a sluice are set to some color a', 
% an instruction to do so must have been given since the last instruction to the 
% traffic lights. In case the lights go to red, this can also be caused by 
% activating an emergency or stopping a gate.

forall l:Lock, s:StreamSide, o:Orientation, a:SingleLight.
  [true*.
   LeavingTrafficLightActuator(l,s,o,a).
   !LeavingTrafficLightCommand(l,s,green)*.
   LeavingTrafficLightActuator(l,s,o,green)]false
&&
forall l:Lock, s:StreamSide, o:Orientation, a:SingleLight.
  [true*.
   LeavingTrafficLightActuator(l,s,o,a).
   !(LeavingTrafficLightCommand(l,s,red)||EmergencyLockCommand(l,activate)||GateCommand(l, s, command_stop))*.
   LeavingTrafficLightActuator(l,s,o,red)]false
&&
forall l:Lock, s:StreamSide.
  [!LeavingTrafficLightCommand(l,s,green)*.
   exists o:Orientation.LeavingTrafficLightActuator(l,s,o,green)]false
&&
forall l:Lock, s:StreamSide.
  [!(LeavingTrafficLightCommand(l,s,red)||EmergencyLockCommand(l,activate)||GateCommand(l, s, command_stop))*.
  exists o:Orientation.LeavingTrafficLightActuator(l,s,o,red)]false
\end{lstlisting}
\subsubsection{Requirement~\ref{causreq31}: Setting the lights of the barrier}
\begin{lstlisting}[language=mCRL2]
% If the traffic lights at the barrier are set to green, a command to do so
% must have been given. If they are set to red, this is due to a similar command,
% or due to an emergency or a stop command for the barrier.

forall s:StreamSide.
  [true*.
   BarrierTrafficLightCommand(s,red).
   !BarrierTrafficLightCommand(s,green)*.
   exists o:Orientation.BarrierTrafficLightActuator(s,o,green)]false
&&
forall s:StreamSide.
  [!BarrierTrafficLightCommand(s,green)*.
   exists o:Orientation.BarrierTrafficLightActuator(s,o,green)]false
&&
forall s:StreamSide.
  [true*.
   BarrierTrafficLightCommand(s,green).
   !(BarrierTrafficLightCommand(s,red)||EmergencyBarrierCommand(activate)||
       BarrierCommand(command_stop))*.
   exists o:Orientation.BarrierTrafficLightActuator(s,o,red)]false
&&
forall s:StreamSide.
  [!(BarrierTrafficLightCommand(s,red)||EmergencyBarrierCommand(activate)||
       BarrierCommand(command_stop))*.
   exists o:Orientation.BarrierTrafficLightActuator(s,o,red)]false
\end{lstlisting}
\subsubsection{Requirement~\ref{causreq32}: Opening the barrier}
\begin{lstlisting}[language=mCRL2]
% If the barrier is opened, this has to be preceded by an instruction to open the barrier.

forall c:ConsoleCommand.(val(c!=command_open) =>
  [true*.
   BarrierCommand(c).
   !BarrierCommand(command_open)*.
   BarrierActuator(do_open)]false)
&&
  [!BarrierCommand(command_open)*.
   BarrierActuator(do_open)]false
\end{lstlisting}
\subsubsection{Requirement~\ref{causreq33}: Closing the barrier}
\begin{lstlisting}[language=mCRL2]
% If the barrier is closed, this has to be preceded by an instruction to close
% the barrier.

forall c:ConsoleCommand.(val(c!=command_close) =>
  [true*.
   BarrierCommand(c).
   !BarrierCommand(command_close)*.
   BarrierActuator(do_close)]false)
&&
  [!BarrierCommand(command_close)*.
   BarrierActuator(do_close)]false
\end{lstlisting}
\subsubsection{Requirement~\ref{causreq34}: Stopping the barrier}
\begin{lstlisting}[language=mCRL2]
% If an emergency stop of the barrier takes place, this has to be preceded by 
% an emergency stop of the barrier, or the instruction to stop that barrier.

forall c:ConsoleCommand.(val(c!=command_stop) =>
  [true*.
   (BarrierCommand(c)||EmergencyBarrierCommand(deactivate)).
   !(BarrierCommand(command_stop)||EmergencyBarrierCommand(activate))*.
   BarrierActuator(do_emergencyStop)]false)
&&
  [!(BarrierCommand(command_stop)||EmergencyBarrierCommand(activate))*.
   BarrierActuator(do_emergencyStop)]false
\end{lstlisting}
\subsection{Operator requirements}
The operator requirements express that if an operator gives an instruction, this will be 
carried out within the lock complex, provided the conditions for execution are right.
The translation of these formulas to modal logic is more involved, and some translations
even become quite big. But the all have the following structure
\begin{lstlisting}[language=mCRL2]
nu X(v:D=initial_d, ...).
     [a]X(new_value_d) &&
     [!a]X(v) && 
     (val(condition) => [b]mu Y.[!c]Y && mu Q.<true>true)
\end{lstlisting}
The maximal fixed point operator \formula{nu} around the modal variable \formula{X}
is used to maintain the observation variable \formula{v}. This variable is of sort \formula{D} and
has initial value \formula{initial\_d}. Whenever an action \formula{a} happens, it gets a new 
value, \formula{new\_value\_d}. Whenever another action happens (\formula{!a}) the value of
\formula{v} remains unchanged. In the translations below up to 15 of such observational
variables are used simultaneously. 

When an action \formula{b} happens, which is in these formulas represent a command by the operator, and the conditions
are right, expressed by \formula{condition}, which is an expression using the observation 
variables, the required effect in the form of action \formula{c} must happen within the 
lock complex. This is expressed using the standard pattern of the modality 
\formula{mu Y.[!c]Y\&\&<true>true} where \formula{mu} is a minimal fixed point operator. 
We use \formula{mu Q....} in the modal formula, to make the symbolic modal formula prover
faster, but it contains no meaning in the formula and can be left out at the cost of
having a slower verification. 

\subsubsection{Requirement~\ref{commandreq1}\label{transcommandreq1}: Close command for the barrier}
Using the three boolean observation variables \formula{downstream\_light\_red}, \formula{upstream\_light\_red}
and \formula{emergency\_mode} it is maintained in the maximal fixed point part of the formula
whether the traffic lights at the downstream side, respectively the upstream side, have been instructed to 
go to red. Using the observation variable \formula{emergency\_mode} it is maintained whether the barrier
is in emergency mode. If the command \formula{command\_close} for the barrier is given under the condition
that the lights are red, and there is no emergency, then within a finite number of actions the instruction
\formula{do\_close} is sent to the barrier actuator. 
\begin{lstlisting}[language=mCRL2]
% If the barrier is instructed to close, it will close provided the traffic lights
% show red and the barrier is not in emergency mode. 

nu X(downstream_light_red:Bool=false,
     upstream_light_red:Bool=false,
     emergency_mode:Bool=false).
 (forall s:StreamSide,sl:SingleLight.[BarrierTrafficLightCommand(s,sl)]
     X(if(s==downstream,sl==red,downstream_light_red),
       if(s==upstream,sl==red,upstream_light_red),
       emergency_mode))&&
 (forall c:EmergencyCommand.[EmergencyBarrierCommand(c)]
     X(downstream_light_red,upstream_light_red,c==activate))&&
 [!((exists s:StreamSide,sl:SingleLight.BarrierTrafficLightCommand(s,sl))||
    (exists c:EmergencyCommand.EmergencyBarrierCommand(c)))
 ]X(downstream_light_red,upstream_light_red,emergency_mode)&&
 (val(downstream_light_red&&upstream_light_red&&!emergency_mode) => 
       [BarrierCommand(command_close)]
       mu Y.[!(BarrierActuator(do_close))]Y&&mu Q.<true>true)
\end{lstlisting}
\subsubsection{Requirement~\ref{commandreq2}: Open command for the barrier}
The structure of this formula is similar to formula~\ref{transcommandreq1}. 
It is slightly more complex as it takes into account that if the barrier lights do not show red,
the barrier does not need to open. This is expressed at lines 19-21 where either the barrier can be 
instructed to open or the traffic lights are measured not to show red. 

\begin{lstlisting}[language=mCRL2]
% If the barrier is instructed to open, it will open provided the traffic lights 
% are set to red, the barrier is not in emergency mode, and the traffic lights 
% show red. 

nu X(downstream_light_red:Bool=false,
     upstream_light_red:Bool=false,
     emergency_mode:Bool=false).
 (forall s:StreamSide,sl:SingleLight.[BarrierTrafficLightCommand(s,sl)]
     X(if(s==downstream,sl==red,downstream_light_red),
       if(s==upstream,sl==red,upstream_light_red),
       emergency_mode))&&
 (forall c:EmergencyCommand.[EmergencyBarrierCommand(c)]
     X(downstream_light_red,upstream_light_red,c==activate))&&
 [!((exists s:StreamSide,sl:SingleLight.BarrierTrafficLightCommand(s,sl))||
    (exists c:EmergencyCommand.EmergencyBarrierCommand(c)))
 ]X(downstream_light_red,upstream_light_red,emergency_mode)&&
 (val(downstream_light_red&&upstream_light_red&&!emergency_mode) => 
    [BarrierCommand(command_open)]
    mu Y.[!(BarrierActuator(do_open)||
            (exists s:StreamSide,o:Orientation,sl:SingleLightStatus.
               val(sl!=show(red))&&BarrierTrafficLightSensor(s,o,sl)))]Y&&
         mu Q.<true>true)
\end{lstlisting}
\subsubsection{Requirement~\ref{commandreq3}\label{transcommandreq3}: Stop command for the barrier}
This formula can be translated without observation variables. It says that whenever the console
command \formula{command\_stop} for the barrier is given, the instruction \formula{do\_emergencyStop} will be
sent to the actuator of the barrier engine, and the the traffic lights will be set to red. 
\begin{lstlisting}[language=mCRL2]
% If the barrier is instructed to stop, it will stop and all four traffic lights
% will go to red.

[true*.BarrierCommand(command_stop)]
  ((mu Y.[!(BarrierActuator(do_emergencyStop))]Y&&mu Q.<true>true)&&
   forall s:StreamSide,o:Orientation.
      mu Z.[!(BarrierTrafficLightActuator(s,o,red))]Z&& mu Q.<true>true)

\end{lstlisting}
\subsubsection{Requirement~\ref{commandreq4}: Emergency command for the barrier}
This translation is comparable to the one in~\ref{transcommandreq3}.
\begin{lstlisting}[language=mCRL2]
% If the barrier receives an emergency instruction, the engines are stopped and
% all traffic lights of the barrier are switched to red. 

[true*.EmergencyBarrierCommand(activate)]
  ((mu Y.[!(BarrierActuator(do_emergencyStop))]Y&&mu Q.<true>true)&&
   forall s:StreamSide,o:Orientation.
      mu Z.[!(BarrierTrafficLightActuator(s,o,red))]Z&& mu Q.<true>true)
\end{lstlisting}
\subsubsection{Requirement~\ref{commandreq5}: Lights command for the barrier}
The requirement says that if a console command says that the lights of the barrier must go to a certain
colour, then the traffic light actuators will receive a conforming command from the controller.
For red this is straightforward, see lines 9-11. 

For green this is much more involved. We need a maximum fixed point variable with five observation
variables, see line 13-17. 
Furthermore, it must be known that the barrier is open, stored in \formula{barrier\_open}. 
But it is only open when the barrier sensor
measured that it was open, while the barrier was opening, stored in the observation variable
\formula{barrier\_opening}. Otherwise, this reading of the barrier sensor
may be meaningless and should be ignored. The barrier is opening when it received an open command
while the traffic lights were red while not in emergency mode, maintained by 
the observation variables at lines 13-15. Keeping track of the values for the observation 
variables on the basis of the inputs and outputs of the lock controller requires the larger part of the
formula.

In lines 60-63 the liveness aspect of this formula is given. If the barrier is open and a command to set
traffic lights to green is received, then both traffic lights will be set to green. 

\begin{lstlisting}[language=mCRL2]
% If the barrier is instructed to show a certain aspect for its traffic lights
% at a stream side, this light is shown, except that for green the barrier must
% be open. The barrier is open when it is sensed to be open, after a valid 
% open command has been given and the barrier did not receive another command 
% afterwards. An open command is valid if it is given when all lights are set to 
% red and the barrier is not in emergency mode, and it remains valid as long as all 
% measurements of the lights indicate that the lights show red. 

(forall s:StreamSide.[true*.BarrierTrafficLightCommand(s,red)]
    forall o:Orientation.mu Y.[!(BarrierTrafficLightActuator(s,o,red))]Y&&
                         mu Q.<true>true) &&
forall s:StreamSide.
 nu X(downstream_light_red:Bool=false,
      upstream_light_red:Bool=false,
      emergency_mode:Bool=false,
      barrier_opening:Bool=false,
      barrier_open:Bool=false).
  (forall s:StreamSide,sl:SingleLight.[BarrierTrafficLightCommand(s,sl)]
      X(if(s==downstream,sl==red,downstream_light_red),
        if(s==upstream,sl==red,upstream_light_red),
        emergency_mode,
        barrier_opening,
        barrier_open))&&
  (forall c:EmergencyCommand.[EmergencyBarrierCommand(c)]
      X(downstream_light_red,
        upstream_light_red,
        c==activate,
        barrier_opening,
        barrier_open))&&
  (forall c:ConsoleCommand.[BarrierCommand(c)]
      X(downstream_light_red,
        upstream_light_red,
        emergency_mode,
        c==command_open&&downstream_light_red&&upstream_light_red&&!emergency_mode, 
        false))&&
  (forall s:StreamSide,o:Orientation,sl:SingleLightStatus.
      [BarrierTrafficLightSensor(s,o,sl)]
         X(downstream_light_red,
           upstream_light_red,
           emergency_mode,
           barrier_opening&&sl==show(red),
           barrier_open))&&
  (forall sp:SensorPosition.[BarrierSensor(sp)]
      X(downstream_light_red,
        upstream_light_red,
        emergency_mode,
        barrier_opening,
        barrier_opening&&sp==sense_open))&&
  [!((exists s:StreamSide,sl:SingleLight.BarrierTrafficLightCommand(s,sl))||
     (exists c:EmergencyCommand.EmergencyBarrierCommand(c))||
     (exists c:ConsoleCommand.BarrierCommand(c))||
     (exists s:StreamSide,o:Orientation,sl:SingleLightStatus.
        BarrierTrafficLightSensor(s,o,sl))||
     (exists sp:SensorPosition.BarrierSensor(sp)))
  ]X(downstream_light_red,
     upstream_light_red,
     emergency_mode,
     barrier_opening,
     barrier_open)&&
  (val(barrier_open)=>
    [BarrierTrafficLightCommand(s,green)]
        forall o:Orientation.mu Y.[!(BarrierTrafficLightActuator(s,o,green))]Y&&
                             mu Q.<true>true)
\end{lstlisting}
\subsubsection{Requirement~\ref{commandreq6}\label{transcommandreq6}: Close command for gates}
By the formulation of the requirement it is clear that we have to maintain the status of
the leaving and entering traffic light for each stream side of each lock, and we need to
maintain whether the lock is in emergency mode. This is done using the observation 
variables at lines 6-8. Then at line 19 it is stated that if the lights are red and there 
is no emergency, if there is a command to close the gates (line 20), the gates will be closed
(lines 21 and 22).
\begin{lstlisting}[language=mCRL2]
% When the operator gives the command to close the gates, the gates will be closed 
% provided the lock is not in emergency mode and the entering and leaving traffic 
% lights are set to red. 

forall l:Lock, s:StreamSide.
  nu X(leaving_light_red:Bool=false,
       entering_light_red:Bool=false,
       emergency_mode:Bool=false).
   (forall dl:DoubleLight.[EnteringTrafficLightCommand(l,s,dl)]
       X(leaving_light_red,dl in {single_red,redred},emergency_mode))&&
   (forall sl:SingleLight.[LeavingTrafficLightCommand(l,s,sl)]
       X(sl==red,entering_light_red,emergency_mode))&&
   (forall c:EmergencyCommand.[EmergencyLockCommand(l,c)]
       X(leaving_light_red,entering_light_red,c==activate))&&
   [!((exists c:EmergencyCommand.EmergencyLockCommand(l,c))||
      (exists dl:DoubleLight.EnteringTrafficLightCommand(l,s,dl))||
      (exists sl:SingleLight.LeavingTrafficLightCommand(l,s,sl)))
   ]X(leaving_light_red,entering_light_red,emergency_mode)&&
   (val(leaving_light_red&&entering_light_red&&!emergency_mode) => 
     [GateCommand(l,s,command_close)]
       forall o:Orientation.mu Y.([!GateActuator(l,s,o,do_close)]Y&&
                            mu Q.<true>true))

     
\end{lstlisting}
\subsubsection{Requirement~\ref{commandreq7}: Open command for gates}
The translation of this property is very similar to the translation in Section~\ref{transcommandreq6} above.
But as the conditions for opening the gates are more complex, we need 12 observation variables
to record the situation at the lock. 
In essence the formula says that if the command is received to open a pair of gates (line 134)
this will happen (lines 136-143). But for this to happen the condition at lines 131-132 must be satisfied.
This condition says that the opposite gates and paddles must be closed, the leaving and entering lights are
red, the water levels around the gates are equal and the lock is not in emergency mode. 
Furthermore, at lines 136-141 it is stated that the gates will receive a command to close the doors, unless
it is measured that the entering or traffic lights happen to show green, due to a malfunction, in which case
it is not necessary for the doors to receive a close command. 

In order to know that the opposite gates are closed, they must have been measured to be closed while closing.
The gates are closing after a valid close command is received. A close command to a pair of gates is valid
if the surrounding traffic lights are red and the lock is not in emergency mode. 

\lstinputlisting[language=mCRL2]{model/requirements/command_gates_open.mcf}
\subsubsection{Requirement~\ref{commandreq8}: Stop command for gates}
As there are no conditions on forwarding a stop command to the gate actuators, this translation
is straightforward. Whenever a stop command of a pair of gates is received, both gates receive that
stop command, and the surrounding traffic lights go to red.
\begin{lstlisting}[language=mCRL2]
% When the operator gives the command to stop the gates, the gates will receive a 
% stop command and the entering and leaving traffic lights are set to red.

forall l:Lock, s:StreamSide.
  [true*.GateCommand(l,s,command_stop)]
     forall o:Orientation.
       (mu Y.([!GateActuator(l,s,o,do_emergencyStop)]Y&&mu Q.<true>true)&&
        mu Y.([!LeavingTrafficLightActuator(l,s,o,red)]Y&&mu Q.<true>true)&&
        mu Y.([!(EnteringTrafficLightActuator(l,s,o,single_red)||
                    EnteringTrafficLightActuator(l,s,o,redred))]Y&&
        mu Q.<true>true))
\end{lstlisting}
\subsubsection{Requirement~\ref{commandreq9}: Close command for paddles}
The close command for paddles is always forwarded to the paddles, except if the lock is in
emergency mode. This gives rise to the following translation with one observation variable
\formula{emergency\_mode}, see line 5.
\begin{lstlisting}[language=mCRL2]
% When the operator gives the command to close paddles, the paddles will be closed, 
% if the lock is not in emergency mode.

forall l:Lock, s:StreamSide.
 nu X(emergency_mode:Bool=false).
  (forall c:EmergencyCommand.[EmergencyLockCommand(l,c)]X(c==activate))&&
  [!(exists c:EmergencyCommand.EmergencyLockCommand(l,c))]X(emergency_mode)&&
  (val(!emergency_mode) => 
    [PaddleCommand(l,s,command_close)]
      forall o:Orientation.mu Y.([!PaddleActuator(l,s,o,do_close)]Y&&
                           mu Q.<true>true))

     
\end{lstlisting}
\subsubsection{Requirement~\ref{commandreq10}: Open command for paddles}
The translation of the requirement that an open command of the paddles is forwarded to the paddle 
actuators is quite complex, because the paddles are only allowed to open if the opposing paddles
and gates are known to be closed and the lock is not in emergency mode, see lines 103-107. The complexity
of the formula is in establishing whether the opposite paddles and gates are closed. These are closed
when they were measured to be closed when closing. The opposite paddles are closing when a close command has been
given when the lock is not in emergency mode. The opposite gates are closing when they received a close command
while its traffic lights are red and the lock is not in emergency mode. The proper closure of the opposite gates
is considered to be non proper when it is measured that its traffic lights do not show red or malfunction.

\lstinputlisting[language=mCRL2]{model/requirements/command_paddles_open.mcf}
\subsubsection{Requirement~\ref{commandreq11}: Stop command for paddles}
Formalising the stop command for paddles is straightforward. Whenever a stop command
for paddles is received (line 5) the paddle actuators will both receive an emergency stop 
(lines 6 and 7).
\begin{lstlisting}[language=mCRL2]
% When the operator gives the command to stop the paddles,
% the paddle actuators will be instructed to stop.

forall l:Lock, s:StreamSide.
  [true*.PaddleCommand(l,s,command_stop)]
    forall o:Orientation.mu Y.([!PaddleActuator(l,s,o,do_emergencyStop)]Y&&
                         mu Q.<true>true)
\end{lstlisting}
\subsubsection{Requirement~\ref{commandreq12}: Emergency command for a lock}
Whenever an emergency activate command from the operator is received (line 6), the
paddle actuators are stopped (line 8), the gate actuators are stopped (line 9), the entering traffic
lights are set to red or red-red (lines 10-12) and the leaving lights are set to red (line 13).
\begin{lstlisting}[language=mCRL2]
% If a lock receives an emergency instruction, the engines of the gates and paddles 
% are stopped and all entering and leaving traffic lights of the lock are switched 
% to red or redred. 

forall l:Lock.
  [true*.EmergencyLockCommand(l,activate)]
    forall s:StreamSide,o:Orientation.
      ((mu Y1.[!PaddleActuator(l,s,o,do_emergencyStop)]Y1&&mu Q.<true>true)&&
       (mu Y2.[!GateActuator(l,s,o,do_emergencyStop)]Y2&&mu Q.<true>true)&&
       (mu Y3.[!(EnteringTrafficLightActuator(l,s,o,single_red)||
                 EnteringTrafficLightActuator(l,s,o,redred))]Y3&&
           mu Q.<true>true)&&
       (mu Y4.[!LeavingTrafficLightActuator(l,s,o,red)]Y4&& mu Q.<true>true))
\end{lstlisting}
\subsubsection{Requirement~\ref{commandreq13}\label{transcommandreq13}: Leaving light commands for a lock}
The translation of this property is amazingly large. Setting a traffic light to red is straightforward,
see lines 10-12 as the command to set a traffic light to red is simply forwarded to the traffic
light.

Setting a traffic light to green is quite another matter. The essential requirement is translated
at lines 188-195. It says that if both gates are open (line 188, left), the entering light shows red
(line 188, right), and a command is received to set a leaving traffic light to green
(line 189), then the actuators of the traffic lights will be instructed to go to 
green (line 191) unless the entering traffic lights fail or
are measured to show green (lines 192-194). 

The complexity lies in determining whether the gates with the leaving light are open. The
gates are open when they are measured to be open while opening. They are opening if they
received the command to open while the entering and leaving lights were red, the water level
was equal, the opposite gates and paddles were closed and the gate is not in emergency mode
(lines 108-110). In order to determine this, a similarly complex check must be made to
establish that the opposite gates and paddles are closed.

\lstinputlisting[language=mCRL2]{model/requirements/command_leaving_lights.mcf}
\subsubsection{Requirement~\ref{commandreq14}: Entering light commands for a lock}
The liveness property that the entering lights are instructed to go to a certain light if 
the operator instructs the lights to do so provided the circumstances allow this, is translated
to a huge modal formula. Its structure is the same as~\ref{transcommandreq13} but it has even
more situations to take into account for which it uses 15 observation variables and requires
more than 200 lines, making it more than half the size of the specification. 

The formula is split in three parts. In lines 12-17 the simple case where the lights
are commanded to go to red or red-red is given, which is always forwarded to the traffic light
actuators. 

In lines 19-28 the situation is described where the entering traffic light is commanded to go 
to red-green. This will be forwarded to the actuator providing the corresponding leaving light 
has been instructed to show red. 

The most complex is the situation when the operator instructs an entering light to go to green.
This can only happen if the corresponding leaving lights are red, which is easy, and if the
gates are open (line 208). Determining whether the gates are open requires that they were measured
to be open while properly opening. This last part requires all the other observational variables and is
quite complex but it follows the same structure as
the translation of for instance~\ref{transcommandreq13}.
\lstinputlisting[language=mCRL2]{model/requirements/command_entering_lights.mcf}

\subsection{Liveness requirements}
\subsubsection{Requirement~\ref{livereq1}\label{translivereq1}: The barrier can always be closed}
This formula says that at any moment during the lifetime of the software controller
a finite sequence can be done leading to sending \formula{do\_close} to the barrier, see line 35. 
This finite sequence only contains allowed actions, namely giving the close command (line 8), setting traffic lights to red (line 9),
deactivate the barrier (line 10), send instructions to the actuators of the gates, paddles, barrier,
and traffic lights (lines 11-21), doing the meaningless skip action (line 22) and reading sensors.

Reading sensors is encoded using the following construct, used a number of times at lines 23-34.
\begin{lstlisting}[language=mCRL2]
exists sv:SensedValue.<a(sv)>true &&
     [exists sv:SensedValue.a(sv)]X
\end{lstlisting}
This says that is possible for an action \formula{a}, representing reading a sensor, to happen with some argument \formula{sv} (line 1), and
if action \formula{a} happens with whatever argument \formula{sv}, the subformula \formula{X} must hold,
meaning that from that point onwards, the barrier can still be closed. This says that 
whatever value \formula{sv} is read from a sensor there is a sequence of allowed actions such that the barrier will be closed.
Or formulated differently, being able to close the barrier does not depend on any particular value that is read from 
a sensor. 
\lstinputlisting[language=mCRL2]{model/requirements/barrier_can_be_closed.mcf}
\subsubsection{Requirement~\ref{livereq2}: Gates can always be closed}
This formula follows the structure of formula~\ref{translivereq1}. 
In essence it says that at any time there is a finite sequence of actions allowing any gate to close
(line 37). This sequence of actions only needs to contain actions for the operator to give the close
command (line 8), setting traffic lights to red (lines 9-11), deactivate an emergency (line 12), operate
the actuators of gates, paddles, the barrier, and traffic lights (lines 13-23), doing skip (line 24) 
and reading sensors. The sensors use the same construct as in the translation of requirement
\ref{translivereq1} such that the concrete measured sensor value does not affect the possibility
to close the gates (lines 25-36).
\lstinputlisting[language=mCRL2]{model/requirements/gate_can_be_closed.mcf}
\subsubsection{Requirement~\ref{livereq3}: Ships can pass}
The formula is translated in the same style as Requirement~\ref{translivereq1}. 
The essence is encoded in two observation variables \formula{east\_door\_open} and \formula{west\_door\_open}
reflecting the status of the doors at a stream side of a lock (line 16 and 17). 
There are two auxiliary observation variables to record whether the water level is equal
(line 18) and whether an equal water level measure can be considered valid (line 19), which concretely means
that the paddles must be open. 

At any time (line 14) it
must be possible to execute a sequence of actions such that the variables \formula{east\_door\_open} and \formula{west\_door\_open} are set to
true (line 101). This indicates that a ship can pass those gates. As this holds for all gates,
the operator can grant any ship passage through the locks at any time.

The operator can carry out a number of operations to let a ship pass. 
He can give the command to close the opposite gates (line 20), open the current gate (line 21),
close the opposite paddles (line 22), open and close the paddles at the gate (lines 93-96),
set the lights around the gates to red (lines 23 and 24), 
and deactivate the emergency status of the lock (line 25). 
Furthermore, the controller is allowed to operate all actuators of the lock complex (lines 26-37)
and do the meaningless action skip (line 38). At lines 97-100 the actuator commands to open 
the gates are given, and this is recorded in the two observation variables. 

For the sensors, we would like, as in the translations of requirements~\ref{translivereq1}
that progress in opening the gates does not depend on the exact values that the sensors report.
However, this is not the case. If the traffic lights, after being set to red, indicate that they are
not red, the gates cannot move. So, it is essential that the traffic lights report back at some
point that they indeed show red. This is encoded at lines 50-58 for the entering lights and at lines
59-66 for the leaving lights. For the entering traffic lights it says that it is possible to do a light measurement
such that the operator and controller are one step closer to opening the gates (lines 51-53), and
if the measurement is not not single red, or not around the gates (line 54), then the measured result
does not matter, and the controller/operator must also be closer to opening the gates for any other
value that the entering traffic lights report. 

At lines 67-74 it is indicated that it is essential to measure that the gates at the other side of the stream
of the lock are closed, as without being able to do so, it is not possible to instruct the gates at this side
to open. Similarly, the paddles at the other side of the stream must be measured to be closed. 
Finally, at lines 83-92 it is indicated that it is essential to measure that the water level around the gates
are equal, as otherwise the gates can also not be constructed to open.

\lstinputlisting[language=mCRL2]{model/requirements/ship_can_pass.mcf}
\subsubsection{Requirement~\ref{livereq4}\label{translivereq4}: Stopping gates prematurely}
This translation is pretty straightforward. Using the observation variable \formula{red\_green}
it is maintained whether the last actuator command to the entering light is to go to red-green.
If so, if a command is given to stop the gate, the entering traffic light must be set to single red.
\begin{lstlisting}[language=mCRL2]
% The entering lights will go to red if opening the gates is stopped prematurely 
% while the lights are supposed to show red-green.

forall l:Lock, s:StreamSide, o:Orientation.
nu X(red_green:Bool=false).
   (forall dl:DoubleLight.[EnteringTrafficLightActuator(l,s,o,dl)]
       X(dl==redgreen)) &&
   [!exists dl:DoubleLight.EnteringTrafficLightActuator(l,s,o,dl)]X(red_green) &&
   (val(red_green) => [GateCommand(l,s,command_stop)]
       mu Y.([!EnteringTrafficLightActuator(l,s,o,single_red)]Y&&
             mu Q.<true>true))
\end{lstlisting}
\subsubsection{Requirement~\ref{livereq5}: Emergency stop of the lock}
This translation is straightforward, cf.~\ref{translivereq4}. The case for the
entering traffic lights is given at lines 5-12 and the leaving lights are treated at lines
14 and 15. 
\lstinputlisting[language=mCRL2]{model/requirements/lock_emergency_to_red.mcf}
\subsubsection{Requirement~\ref{livereq6}: Emergency stop of the barrier}
Like requirement~\ref{translivereq4}, the translation of this requirement is straightforward. 
\begin{lstlisting}[language=mCRL2]
% The traffic lights will go to red if the barrier gets an emergency. 

[true*.EmergencyBarrierCommand(activate)]
   forall s:StreamSide,o:Orientation.
      mu Y.([!BarrierTrafficLightActuator(s,o,red)]Y&&mu Q.<true>true)
\end{lstlisting}

\section{The verification scripts}
\label{app:script}
The first shell script below indicates how the requirements are proven. The first parameter (\formula{\$1}) 
indicates the name of the requirement, and the second parameter (\formula{\$2}) is used to indicate
the memory in Gbyte used by the tool \formula{pbessolvesymbolic}. This second parameter
ranges from 4 to 64GByte. 
\lstinputlisting[language=tools]{model/runreq}
The second shell script indicates how the state space of the model is generated using the symbolic
state space generator \formula{lpsreach}.
\begin{lstlisting}[language=tools]
mcrl22lps marijkesluizen_wfaults_spec.mcrl2 scratch/temp.lps -v &&
lpsfununfold scratch/temp.lps -v | lpssuminst | \
  lpsparunfold -s"LockStreamSideOrientationTriple" | \
  lpsparunfold -s"LockStreamSideTuple" | lpsrewr -v > scratch/temp2.lps &&
lpsreach --memory-limit=4 scratch/temp2.lps  -v \
  --cached -rjittyc --groups=used --chaining --saturation --timings
\end{lstlisting}
\end{document}